\documentclass[aps,preprint,onecolumn,11pt]{revtex4}
\usepackage{hyperref}
\usepackage{amsfonts}
\usepackage{amsmath}

\usepackage{amssymb}
\usepackage{tensor}
\usepackage{graphicx}%
\usepackage{bigints}
\usepackage{graphicx}
\usepackage{subfigure}
\usepackage{epsf}
\usepackage{bm}
\usepackage{amsmath}
\usepackage{amsfonts}
\usepackage{amssymb}
\usepackage{graphicx}
\usepackage{tabularx}
\usepackage{color}%
\providecommand{\U}[1]{\protect\rule{.1in}{.1in}}

\newcommand{\ie}{\begin{equation}}
\newcommand{\fe}{\end{equation}}

\newcommand{\mincir}{\raise
-3.truept\hbox{\rlap{\hbox{$\sim$}}\raise4.truept\hbox{$<$}\ }}
\newcommand{\magcir}{\raise
-3.truept\hbox{\rlap{\hbox{$\sim$}}\raise4.truept\hbox{$>$}\ }}

\providecommand{\U}[1]{\protect\rule{.1in}{.1in}}

\usepackage{tikz,xcolor,hyperref}

\definecolor{lime}{HTML}{A6CE39}
\DeclareRobustCommand{\orcidicon}{%
	\begin{tikzpicture}
	\draw[lime, fill=lime] (0,0) 
	circle [radius=0.16] 
	node[white] {{\fontfamily{qag}\selectfont \tiny ID}};
	\draw[white, fill=white] (-0.0625,0.095) 
	circle [radius=0.007];
	\end{tikzpicture}
	\hspace{-2mm}
}

\foreach \x in {A, ..., Z}{%
	\expandafter\xdef\csname orcid\x\endcsname{\noexpand\href{https://orcid.org/\csname orcidauthor\x\endcsname}{\noexpand\orcidicon}}
}

\begin{document}

\title{\Large{Exploring antisymmetric tensor effects on black hole shadows and quasinormal frequencies}}


\author{A. A. Ara\'{u}jo Filho \orcidB{}}
\email{dilto@fisica.ufc.br}


\affiliation{Departamento de Física, Universidade Federal da Paraíba, Caixa Postal 5008, 58051-970, João Pessoa, Paraíba, Brazil}

\author{J. A. A. S. Reis\orcidA{}}
\email{jalfieres@gmail.com}

\affiliation{Universidade Estadual do Sudoeste da Bahia (UESB), Departamento de Ciências Exatas e Naturais, Campus Juvino Oliveira, Itapetinga -- BA, 45700-00, Brazil}


\author{H. Hassanabadi\orcidC{}}
\email{hha1349@gmail.com}

\affiliation{Physics Department, Shahrood University of Technology, Shahrood, Iran}

\affiliation{Department of Physics, University of Hradec Kr$\acute{a}$lov$\acute{e}$, Rokitansk$\acute{e}$ho 62, 500 03 Hradec Kr$\acute{a}$lov$\acute{e}$, Czechia.}


\begin{abstract}

This study explores the impact of antisymmetric tensor effects on spherically symmetric black holes, investigating photon spheres, shadows, emission rate and quasinormal frequencies in relation to a parameter which triggers the Lorentz symmetry breaking. We examine these configurations without and with the presence of a cosmological constant. In the first scenario, the Lorentz violation parameter, denoted as $\lambda$, plays a pivotal role in reducing both the photon sphere and the shadow radius, while also leading to a damping effect on quasinormal frequencies.  Conversely, in the second scenario, as the values of the cosmological constant ($\Lambda$) increase, we observe an expansion in the shadow radius. Also, we provide the constraints of the shadows based on the analysis observational data obtained from the Event Horizon Telescope (EHT) focusing on Sagittarius $A^{*}$ shadow images. Additionally, with the increasing $\Lambda$, the associated gravitational wave frequencies exhibit reduced damping modes.

\end{abstract}
\maketitle


\section{Introduction}

Lorentz symmetry, a fundamental pillar of modern physics, postulates the consistent applicability of physical laws across all inertial reference frames. While firmly established as a fundamental principle, corroborated by extensive experimental and observational background, it has become evident that Lorentz symmetry may exhibit deviations under specific energy conditions within diverse theoretical approaches, such as string theory \cite{1}, loop quantum gravity \cite{2}, Horava--Lifshitz gravity \cite{3}, non--commutative field theory \cite{4}, Einstein--aether theory \cite{5}, massive gravity \cite{6}, $f(T)$ gravity \cite{7}, very special relativity \cite{8}, and more.

Lorentz symmetry breaking (LSB) can be manifested in two distinct manners: explicit and spontaneous \cite{bluhm2006overview}. The first one case occurs when the Lagrangian density lacks Lorentz invariance, resulting in the formulation of distinct physical laws in specific reference frames. Conversely, the second one arises when the Lagrangian density maintains Lorentz invariance, but the ground state of a physical system does not exhibit Lorentz symmetry \cite{bluhm2008spontaneous}.

The investigation of spontaneous Lorentz symmetry breaking \cite{9,10,11,12,13,KhodadiPoDU2023} finds its foundation in the Standard Model Extension. Within this framework, the simplest field theories are encapsulated by bumblebee models \cite{1,10,11,12,13, KhodadiEPJC2023,KhodadiEPJC20232,KhodadiPRD2022,CapozzielloJCAP2023}. In these models, a vector field, termed the bumblebee field, acquires a non--zero vacuum expectation value (VEV). This aspect establishes a distinct direction, resulting in the local Lorentz invariance violation for particles, which in turn leads to remarkable consequences for instance to the thermodynamic properties \cite{petrov2021bouncing2,aaa2021thermodynamics,aa2021lorentz,araujo2021thermodynamic,aa2022particles,reis2021thermal,araujo2021higher,araujo2022thermal,araujo2022does,paperrainbow,anacleto2018lorentz}.

Ref. \cite{14} presents an exact solution for a static and spherically symmetric spacetime within the framework of bumblebee gravity. Analogously, a Schwarzschild--like solution has undergone rigorous examination from multiple perspectives, including Hawking radiation \cite{16}, the accretion process \cite{17,18} gravitational lensing \cite{15}, and quasinormal modes \cite{19}.

Following this, Maluf et al. expanded upon these discoveries by deriving an (A)dS--Schwarzschild--like solution, relaxing the vacuum conditions \cite{20}. Additionally, Xu et al. introduced innovative categories of static spherical bumblebee black holes by incorporating a background bumblebee field with a non--zero temporal component, exploring their thermodynamic characteristics and observational consequences in Refs. \cite{21,22,23,24}.

Ding et al. investigated the domain of rotating bumblebee black holes in Ref. \cite{25,26}, covering a wide range of topics including their accretion processes \cite{28}, shadows \cite{27}, quasi--periodic oscillations \cite{30}, and quasinormal modes \cite{29}. Furthermore, an exact rotating BTZ--like black hole solution was derived in \cite{31}, and its quasinormal modes were subject to analytical investigation in reference \cite{32}.

A Schwarzschild--like black hole incorporating a global monopole was introduced in \cite{33}, and its quasinormal modes were analyzed \cite{34,35}. Moreover, various other black hole solutions were explored within the framework of the Bumblebee gravity model in Refs. \cite{36,37,38}, as well as in the context of metric affine formalism \cite{nascimento2022vacuum,delhom2022radiative,delhom2022spontaneous}. Additionally, a traversable bumblebee wormhole solution was recently proposed in the literature \cite{39}, with subserquents investigations of its corresponging gravitational waves \cite{40,41}.

Beyond the vector field theories, another approach for exploring LSB involves a rank--two antisymmetric tensor field referred to as the Kalb--Ramond field \cite{42,maluf2019antisymmetric}. It naturally arises in the spectrum of bosonic string theory \cite{43}. When this field is non--minimally coupled to gravity and obtains a non--zero vacuum expectation value, it breaks the Lorentz symmetry spontaneously. In Ref. \cite{44}, an exact solution for a static and spherically symmetric configuration within this context is demonstrated. This discovery was followed by an exploration of the behavior of both massive and massless particles near this static spherical Kalb--Ramond black hole in Reference \cite{45}. Additionally, Ref. \cite{46} investigates the gravitational deflection of light and the shadows cast by rotating black holes.

Over the past few years, there has been a significant focus on the exploration of gravitational waves and their spectra, as evidenced by recent studies \cite{bombacigno2023landau,aa2023analysis,boudet2022quasinormal,hassanabadi2023gravitational,amarilo2023gravitational}. This heightened interest can be attributed in large part to the remarkable advancements in gravitational wave detection technology, notably exemplified by the VIRGO and LIGO detectors. These advanced instruments have played a crucial role in providing profound understanding into the captivating domain of black hole physics \cite{abbott2016gw150914,abramovici1992ligo,grishchuk2001gravitational,vagnozzi2022horizon}. In Ref. \cite{yang2023static}, presents novel exact solutions for static and spherically symmetric spacetime, both in the presence and absence of the cosmological constant. These solutions are derived within the context of a non--zero vacuum expectation value background of the Kalb--Ramond field.

Within the field of black hole research, a significant facet of investigation revolves around the exploration of quasinormal modes (QNMs). These QNMs represent intricate oscillation frequencies that emerge as a consequence of black holes responding to initial perturbations. To derive these frequencies, specific boundary conditions must be imposed \cite{berti2009quasinormal,konoplya2011quasinormal}.
The study of gravitational waves and their spectra over the last years \cite{bombacigno2023landau,aa2023analysis,boudet2022quasinormal,hassanabadi2023gravitational,amarilo2023gravitational}, particularly, with the advancements in gravitational wave detectors such as VIRGO and LIGO. These ones have provided valuable insights into black hole physics \cite{abbott2016gw150914,abramovici1992ligo,grishchuk2001gravitational,vagnozzi2022horizon}. One aspect of this research involves investigating the quasinormal modes (QNMs), which are complex oscillation frequencies that arise in the response of black holes to initial perturbations. These frequencies can be obtained under specific boundary conditions \cite{berti2009quasinormal,konoplya2011quasinormal}. 

In this work, we study the impact of anti--symmetric tensor effects, which triggers the Lorentz symmetry breaking, on spherically symmetric black holes, analyzing photon spheres, shadows, and quasinormal frequencies. We explore these configurations both in the absence and presence of a cosmological constant.


\section{The antisymmetric black hole solution}

We commence by considering the Einstein--Hilbert action, which is non-minimally coupled to a self--interacting Kalb--Ramond (KR) field, expressed in the following form \cite{altschul2010lorentz}:
\ie
\begin{split}
\label{action}
S = & \frac{1}{2 \kappa^{2}} \int \mathrm{d}^{4}x \sqrt{-g} \left[ R - 2\Lambda - \frac{1}{6}H^{\mu\nu\rho}H_{\mu\nu\rho} - V(B_{\mu\nu}B^{\mu\nu}) + \xi_{2} B^{\rho\mu}B\indices{^\nu_\mu}  + \xi_{3} B^{\mu\nu}B_{\mu\nu} R        \right] \\
& + \int \mathrm{d}^{4}x\sqrt{-g} \mathcal{L}_{m},
\end{split}
\fe
where $\xi_{2}$ and $\xi_{3}$ are the coupling constants between the Ricci tensor and the Kalb--Ramond field, $\kappa = 8 \pi G$, $\Lambda$ is the cosmological constant, and the field strength $H_{\mu\nu\rho} \equiv \partial_{[\mu}B_{\nu\rho]}$. The potential $V(B_{\mu\nu}B^{\mu\nu})$ is responsible for triggering the spontaneous Lorentz symmetry breaking. With such an aspect, we are able to maintain this theory invariant under local Lorentz transformations.

In order to obtain the gravitational field equations, we vary Eq. (\ref{action}) with respect to the metric tensor $g^{\mu\nu}$, which gives
\ie
R_{\mu\nu} - \frac{1}{2}g_{\mu\nu}R + \Lambda g_{\mu\nu} = T^{m}_{\mu\nu} + T^{KR}_{\mu\nu} = \mathcal{T}_{\mu\nu},
\fe
where $\mathcal{T}_{\mu\nu}$ is the total stress--energy tensor, $T^{m}_{\mu\nu}$ is the stress--energy of the matter fields and $T^{KR}_{\mu\nu}$ being given by
\ie
\begin{split}
T^{KR}_{\mu\nu} = & \frac{1}{2}H_{\mu\alpha\beta}H\indices{_\nu^\alpha
^\beta} - \frac{1}{12}g_{\mu\nu}H^{\alpha\beta\rho}H_{\alpha\beta\rho} + 2 V^{\prime}(X)B_{\alpha\mu}B\indices{^\alpha_\nu} - g_{\mu\nu}V(X) \\
& \xi_{2} \left\{ \frac{1}{2}g_{\mu\nu}B^{\alpha\gamma}B\indices{^\beta_\gamma}R_{\alpha\beta} - B\indices{^\alpha_\mu}B\indices{^\beta_\nu} R_{\alpha\beta} - B^{\alpha\beta}B_{\nu\beta}R_{\mu\alpha} - B^{\alpha\beta}B_{\mu\beta}R_{\nu\alpha} \right. \\
& \left.  \frac{1}{2} \nabla_{\alpha}\nabla_{\mu}(B^{\alpha\beta}B_{\nu\beta})  + \frac{1}{2} \nabla_{\alpha}\nabla_{\nu}(B^{\alpha\beta}B_{\mu\beta}) - \frac{1}{2} \nabla^{\alpha}\nabla_{\alpha}B\indices{_\mu^\gamma}B_{\nu\gamma}  \right. \\
& \left. -\frac{1}{2}g_{\mu\nu} \nabla_{\alpha}\nabla_{\beta}(B^{\alpha\gamma}B\indices{^\beta_\gamma})          \right\}.
\end{split}
\fe
In this context, the prime symbol denotes the derivative with respect to the argument of the respective functions. Utilizing the Bianchi identity, we can ascertain that the $\mathcal{T}_{\mu\nu}$ is conserved. To induce a non-zero vacuum expectation value (VEV) for the Kalb--Ramond field, denoted as \( \langle B_{\mu\nu} \rangle = b_{\mu\nu} \), we consider a potential of general form \( V = V(B_{\mu\nu} B^{\mu\nu} \pm b^2) \). Here, the choice of the sign \( \pm \) ensures that \( b^2 \) is a positive constant. As a result, the vacuum expectation value (VEV) is defined by the condition \( b_{\mu\nu} b^{\mu\nu} = \mp b^2 \). The gauge invariance \( B_{\mu\nu} \rightarrow B_{\mu\nu} + \partial_{[\mu} \Gamma_{\nu]} \) associated with the Kalb--Ramond field is spontaneously broken. Since the non--minimal coupling of the KR field with gravity is present, this symmetry--breaking process due to a VEV background gives rise to a violation of local Lorentz invariance for particles. Additionally, the term \( \xi_3 B_{\mu\nu} B^{\mu\nu} R \) in Eq. (\ref{action}) becomes \( \mp \xi_3 b^2 R \) in the vacuum state, and this can be integrated into the Einstein--Hilbert terms through a suitable redefinition of variables.

For convenience, the antisymmetric tensor \( B_{\mu\nu} \) can be decomposed as \( B_{\mu\nu} = \tilde{E}_{[\mu} v_{\nu]} + \epsilon_{\mu\nu\alpha\beta} v^\alpha \tilde{B}^\beta \), where \( v^\alpha \) is a timelike 4--vector. The pseudo-fields \( \tilde{E}^\mu \) and \( \tilde{B}^\mu \) are spacelike and satisfy the conditions \( \tilde{E}^\mu v_\mu = \tilde{B}^\mu v_\mu = 0 \). Drawing a parallel to Maxwell's electrodynamics, these pseudo-fields can be thought of as pseudo--electric and pseudo--magnetic fields.

If we assume that the only non--zero terms in the VEV are \( b_{10} = -b_{01} = \tilde{E}(r) \), or similarly, \( {\bf{b}}^{2} = -\tilde{E}(r) \, \mathrm{d}t \wedge \mathrm{d}r \), it becomes apparent that the vacuum field turns out to present a configuration of a pseudo--electric field. As a result, this specific configuration leads to the vanishing of the Kalb--Ramond (KR) field strength, meaning \( H_{\lambda\mu\nu} = 0 \) or \( {\bf{H}}^3 = \mathrm{d} {\bf{b}}^{2} = 0 \).

In this study, we concentrate on exploring a static and spherically symmetric spacetime, set against a non--zero (VEV) for the KR field. The proposed metric form is described as follows:
The metric tensor for this spacetime is presented in terms of the following line element:
\ie
\mathrm{d}s^2 = -A(r) \mathrm{d}t^2 + B(r) \mathrm{d}r^2 + r^2 \mathrm{d}\theta^2 + r^2 \sin^2 \theta \mathrm{d}\phi^2.
\fe
Here, \(A(r)\) and \(B(r)\) are functions of the radial coordinate \(r\), which will be determined by the underlying dynamics of the system, including the KR field. Accordingly, the pseudo--electric field \( \tilde{E}(r) \) can be reformulated as $
\tilde{E}(r) = \left| b \right| \sqrt{ \frac{A(r) B(r)}{2} },
$
ensuring that the constant norm condition \( b_{\mu\nu} b^{\mu\nu} = -b^2 \) is fulfilled.

It becomes suitable to rewrite the field equation in terms of $b$ as follows:
\ie
\begin{split}
R_{\mu\nu} =& \,\, \Lambda g_{\mu\nu} + V^{\prime}( b_{\mu\alpha} b\indices{_\nu^\alpha} + b^{2} g_{\mu\nu}) + \xi_{2} \left[ g_{\mu\nu} b^{\alpha\gamma} b^\beta{}_\gamma R_{\alpha\beta} - b\indices{^\alpha_\mu} b\indices{^\beta_\nu} R_{\alpha\beta} - b^{\alpha\beta}b_{\mu\beta}R_{\nu\alpha}   \right. \\ 
& \left. - b^{\alpha\beta} b_{\nu\beta} R_{\mu\alpha}
 + \frac{1}{2} \nabla_\alpha \nabla_\mu (b^{\alpha\beta} b_{\nu\beta}) + \frac{1}{2} \nabla_\alpha \nabla_\nu (b^{\alpha\beta} b_{\mu\beta}) - \frac{1}{2} \nabla^\alpha \nabla_\alpha (b\indices{_\mu^\gamma} b_{\nu\gamma}) \right] \label{modifiedfieldeq}.
\end{split}
\fe
Given the specified metric ansatz, the field equations, denoted as Eq. (\ref{modifiedfieldeq}), can be explicitly reformulated as follows \cite{yang2023static}:
\ie
\begin{split}
& \frac{2A^{\prime}}{A} - \frac{A^{\prime}B^{\prime}}{A B} - \frac{A^{\prime 2}}{A^{2}} + \frac{4 A^{\prime}}{r A} + \frac{4\Lambda}{1-\lambda}B = 0, \\
 & \frac{2A^{\prime\prime}}{A} - \frac{A^{\prime} B^{\prime}}{A B} - \frac{A^{\prime 2}}{A^{2}} - \frac{4 B^{\prime}}{r B} + \frac{4\Lambda}{1-\lambda}B = 0,\\
 & \frac{2 A^{\prime\prime}}{A} - \frac{A^{\prime}B^{\prime}}{AB} - \frac{A^{\prime 2}}{A^{2}} + \frac{1+\lambda}{\lambda r}\left( \frac{A^{\prime}}{A}-\frac{B^{\prime}}{B} \right) - (1 - \Lambda r^{2} - b^{2}r^{2} V^{\prime})\frac{2B}{\lambda r^{2}} + \frac{2(1-\lambda)}{\lambda r^{2}} = 0,
 \end{split}
\fe
where $ \lambda \equiv \frac{\xi_2 b^2}{2}$.


\subsection{For $\Lambda = 0$}

Initially, in the scenario where a cosmological constant is absent, our aim is to formulate a black hole solution resembling that of the Schwarzschild geometry within the framework of this theory. Here, we assume \( V' = 0 \), indicating that the vacuum expectation value (VEV) is situated at a local minimum of the potential. This condition can be conveniently verified, for example, by considering a smooth quadratic potential $V = \frac{1}{2} \sigma X^2$, where $X \equiv B_{\mu\nu} B^{\mu\nu} + b^2$ and $\sigma$ serves as a coupling constant.

After some algebraic manipulations, we obtain \cite{yang2023static}:
\ie
A(r)= \frac{1}{B(r)} = \frac{1}{1-\lambda} - \frac{2M}{r},
\fe
which yields
\ie
\mathrm{d}s^{2} = - \left( \frac{1}{1-\lambda} - \frac{2M}{r}   \right) \mathrm{d}t^{2} + \frac{\mathrm{d}r^{2}}{\frac{1}{1-\lambda} - \frac{2M}{r} } + r^{2}\mathrm{d}\theta^{2} + r^{2} \sin^{2}\mathrm{d}\varphi^{2}.
\fe
It gives rise to the following event horizon
\ie
r_{\Lambda=0} = 2 (1-\lambda) M.
\fe
In the next subsections, we shall calculate the photon spheres as well as the shadows of the black hole under consideration. In addition, the \textit{quasinormal} frequencies are addressed in this context.

\subsubsection{Photon sphere and shadows}

For a comprehensive understanding of particle dynamics and the patterns of light rays in proximity to black holes, it is fundamental to comprehend the aspects of photon spheres. These spherical regions play an integral role in deciphering the shadows projected by black holes, as well as in understanding the influence of the KR field on the spacetime under investigation.

In terms of black hole dynamics, the significance of the photon sphere, i.e., often designated as the critical orbit, necessitates the utilization of the Lagrangian method for instance to compute the null geodesics. The objective of this study is to scrutinize the influence of specific parameters, specifically $M$ and $\lambda$, on it. To offer further clarification, we assert:
\ie
\mathcal{L} = \frac{1}{2} g_{\mu\nu}\Dot{x}^{\mu}\Dot{x}^{\nu}.
\fe
Upon setting the angle to \(\theta = \pi/2\), the aforementioned equation simplifies as follows:
\ie
g_{00}^{-1} E^{2} + g_{11}^{-1} \Dot{r}^{2} + g_{33}L^{2} = 0.\label{sep123aration}
\fe
In this context, \(L\) signifies the angular momentum while \(E\) represents the energy \cite{HouPRD2022}. Consequently, Eq. (\ref{sep123aration}) can be expressed as:
\ie
\Dot{r}^{2} = E^{2} - \left( \frac{1}{1-\lambda }-\frac{2 M}{r} \right)\left(  \frac{L^{2}}{r^{2}} \right),
\fe
where the effective potential is denoted as $\mathcal{V} \equiv \left( \frac{1}{1-\lambda }-\frac{2 M}{r} \right)\left(  \frac{L^{2}}{r^{2}} \right)$. To ascertain the critical radius, one must solve the equation \(\partial \mathcal{V}/\partial r = 0\). This solution reads:
\ie
r_{c} = 3(1- \lambda) M.
\fe
Here, we denote $r_{c}$ is the radius of the critical orbits. Obviously, when $\lambda \rightarrow 0$, we recover the Schwarzschild case. Furthermore, the Lorentz symmetry breaking brings about a constraint to parameter $\lambda$, i.e., bounded from above. In other words, the physical limits of such a value is $\lambda = 1$. After that, $r_{c}$ will not show physical values. Additionally, recent literature, including references \cite{aa2023implications,aa2023analysis}, has explored analogous studies in the realm of dark matter. It is also important to underscore that the appearance of dual photon spheres has been reported within the framework of the Simpson--Visser solution \cite{tsukamoto2021gravitational, tsukamoto2022retrolensing}, and others \cite{guerrero2022multiring}.

\begin{table}[!h]
\begin{center}
\caption{\label{p1h2o3t4onspheres} The critical orbit, denoted by \(r_{c}\), is illustrated for a range of values pertaining to mass \(M\), and parameter \(\lambda\).}
\begin{tabular}{c c c ||| c c c c } 
 \hline\hline
 $M$ & $\lambda$ & $r_{c}$ & $M$ & $\lambda$ & $r_{c}$ &   \\ [0.2ex] 
 \hline 
 1.00 & 0.00 & 3.00 & ----- & ----- & ----- &  \\ 

 1.00 & 0.10 & 2.70 & 1.00 & 0.10 & 2.70 & \\
 
 1.00 & 0.20 & 2.40 & 2.00 & 0.10 & 5.40 &  \\
 
 1.00 & 0.30 & 2.10 & 3.00 & 0.10 & 8.10 &    \\
 
 1.00 & 0.40 & 1.80 & 4.00 & 0.10 & 10.8 &   \\
 
 1.00 & 0.50 & 1.50 & 5.00 & 0.10 & 13.5 &  \\
 
 1.00 & 0.60 & 1.20 & 6.00 & 0.10 & 16.2 &  \\
 
 1.00 & 0.70 & 0.90 & 7.00 & 0.10 & 18.9 &   \\
 
 1.00 & 0.80 & 0.60 & 8.00 & 0.10 & 21.6 &  \\
 
 1.00 & 0.90 & 0.30 & 9.00 & 0.10 & 24.3 &   \\
 
 1.00 & 1.00 & 0.00 & 10.0& 0.10 & 27.0 &   \\[0.2ex] 
 \hline \hline
\end{tabular}
\end{center}
\end{table}

The left side of Table \ref{p1h2o3t4onspheres} presents various calculations of the critical orbit \(r_{c}\), each corresponding to different values of \(\lambda\) while maintaining \(M=1\). In this setting, as the Lorentz--violating parameter \(\lambda\) increases, \(r_{c}\) undergoes a corresponding decrease until it reaches a trivial contribution (when $\lambda=1$). Given that this parameter is intrinsically linked to the Kalb--Ramond field, our results indicate that Lorentz--violating coefficients exert a considerable influence on this scenario by reducing the radius of the photon sphere. On the other hand, when the mass \(M\) is varied, a significant increase in the critical orbits is observed.

The scrutiny of shadows within the framework of the Kalb--Ramond field and black hole configurations is critically important. Characterized by the unique silhouette of a black hole contrasted against a luminous backdrop, these shadows serve as windows into the geometry of spacetime and nearby gravitational interactions. Analysis of these phenomena enables us to extract invaluable information for refining theoretical models, thereby providing a more robust validation of gravitational theories. To facilitate our investigation, we introduce two novel parameters, as outlined below \cite{87,88}:
\ie
\label{newparameters}
\xi  = \frac{L}{E}
\text{ and }
\eta  = \frac{\mathcal{K}}{{E^2}},
\fe
where \(\mathcal{K}\) is commonly referred to as the Carter constant \cite{89,90}. Following a series of algebraic manipulations, the resulting expression is:
\ie
{\xi ^2} + \eta  =  \frac{r_{c}^2}{f(r_{c})}.
\fe

In our quest to determine the radius of the shadow, we will employ the celestial coordinates \(\tilde{\alpha}\) and \(\beta\) \cite{singh2018shadow,campos2022quasinormal,heidari2023gravitational,hassanabadi2023gravitational} as: $\alpha=-\xi$ and $\beta=\pm\sqrt{\eta}$. Thereby, we are able to write the radius shadow as follows
\begin{equation}
\mathcal{R}_{\Lambda=0} = \frac{r_{c}}{{\sqrt{f(r_{c})}}} = \frac{3 (1-\lambda) M}{\sqrt{\frac{1}{3-3 \lambda }}}.
\end{equation}

\begin{figure}
    \centering
    \includegraphics[scale=0.25]{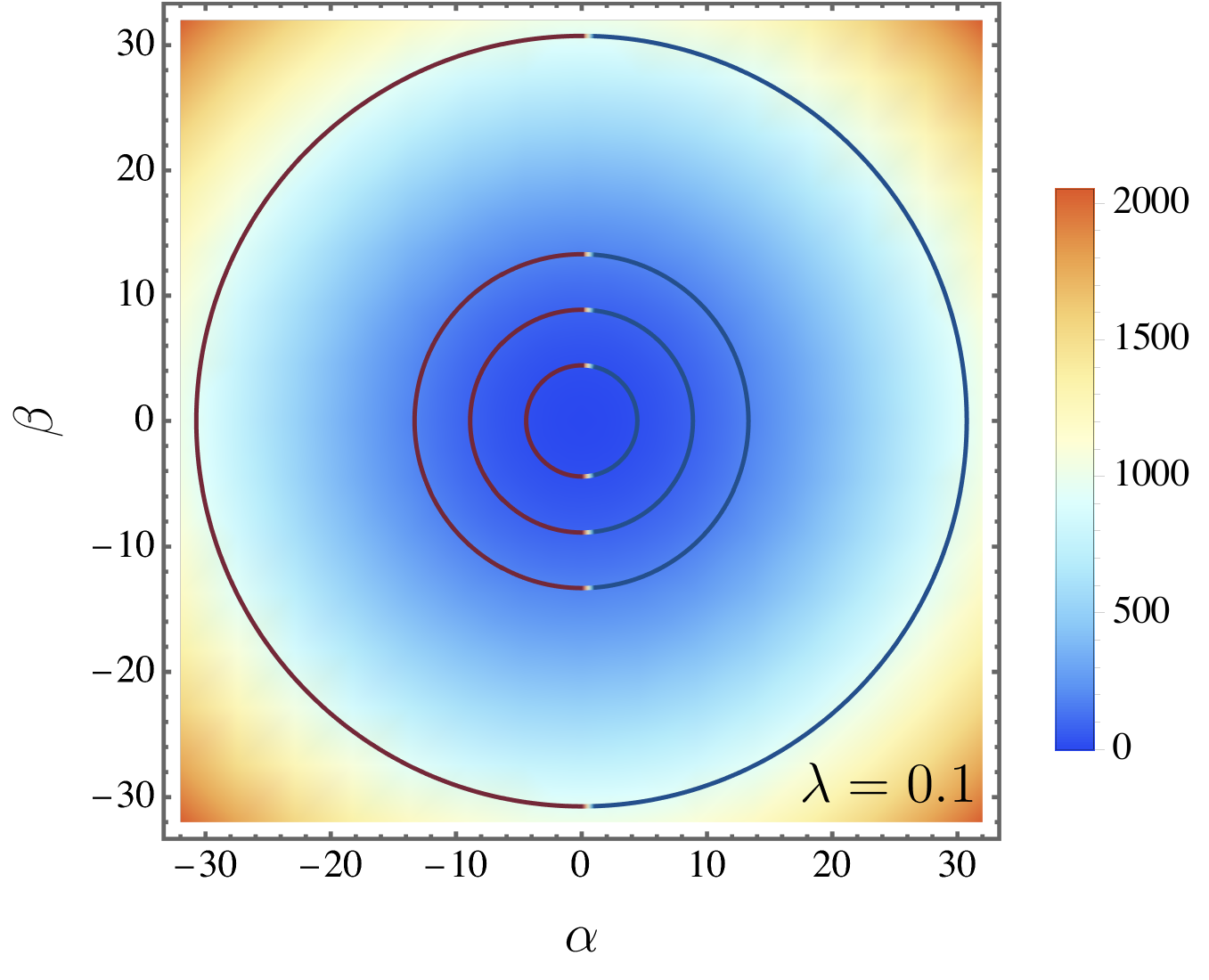}
    \includegraphics[scale=0.25]{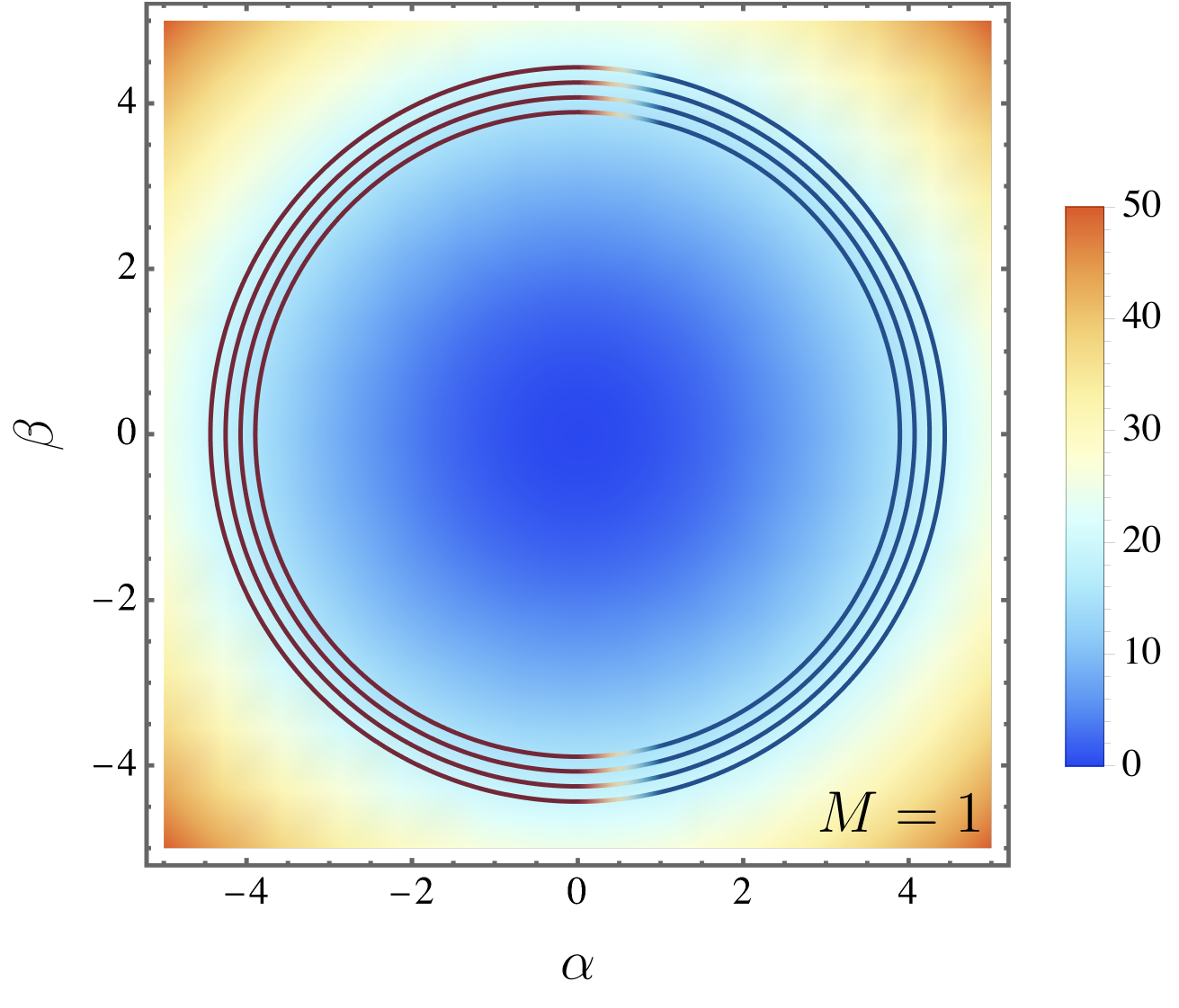}
    \caption{The circles in the figure illustrate shadows formed by varying the parameters \(M\) and \(\lambda\). Specifically, on the left section, the shadows are generated with \(M\) values ranging from 1 (representing the innermost radius) to 4 (representing the outermost radius), focusing solely on integer increments. In contrast, on the right features shadows that vary with \(\lambda\) values from 0.1 (outermost radius) to 0.175 (innermost radius), when \(M\) is set to 1. The radius increments by 0.025 for each successive circle in this case. Color density refers to the magnitude associated with the coordinates $\alpha$ and $\beta$, as indicated in the plot legends situated on the right side of each respective graph.}
    \label{shadows23c45omplete}
\end{figure}

The illustration in Fig. \ref{shadows23c45omplete} demonstrates the impact of parameters $M$ and $\lambda$ on the black hole shadows. Notably, from a broader perspective, an increase in the parameter $\lambda$, i.e., $\lambda=0.100$, $\lambda=0.125$, $\lambda=0.150$, and $\lambda=0.175$, is associated with a decrease in the radius of the shadows, which is shown in the right side. This behavior can be directly attributed to the presence of Lorentz violation in the theoretical framework under consideration.

Furthermore, on the left section of Fig. \ref{shadows23c45omplete} presents black hole shadows for mass values ranging from \(M=1\) (corresponding to the innermost radius) to \(M=4\) (corresponding to the outermost radius), considering only integer increments while keeping \(\lambda\) fixed at $\lambda=0.1$. It is evident that as the mass \(M\) increases, the radius of the shadow expands correspondingly.

\subsubsection{Quasinormal modes}

In the ringdown phase, the phenomenon of \textit{quasinormal} modes \cite{94,95,96,97,98,99,100,101,102,103,104,105,106,107} comes to the fore, exhibiting unique oscillation patterns that remain unaffected by the system's initial state. These modes serve as a fingerprint for it, arising from the inherent vibrations of spacetime and remaining consistent regardless of specific initial conditions. In contrast to \textit{normal} modes, which are associated with closed systems, \textit{quasinormal} modes are relevant to open systems, leading to a gradual energy loss through the emission of gravitational waves. From a mathematical standpoint, these modes are identified as poles in the complex Green's function.

To determine the frequencies of quasinormal modes, it is essential to solve the wave equation in a system governed by the background metric \(g_{\mu\nu}\). However, deriving analytical solutions for these modes is frequently a complex task. A range of methods for solving for these modes has been proposed in the scientific literature, with the WKB (Wentzel--Kramers--Brillouin) method standing out as one of the most widely used. This technique was significantly influenced by the groundbreaking work of Will and Iyer \cite{iyer1987black,iyer1987black1}. Subsequent refinements, extending the method to the sixth order, have been made by Konoplya \cite{konoplya2003quasinormal}. For the computations in our study, we focus on perturbations mediated through the scalar field, utilizing the Klein--Gordon equation within the context of curved spacetime
\ie
\frac{1}{\sqrt{-g}}\partial_{\mu}(g^{\mu\nu}\sqrt{-g}\partial_{\nu}\Phi) = 0.\label{KleinGordon}
\fe
While the exploration of backreaction effects within this context is undoubtedly important, the focus of this manuscript is directed differently. We primarily center our investigation on the scalar field, treating it as a minor perturbation. The inherent spherical symmetry of the scenario at hand allows for a specific decomposition of the scalar field, which we write as follows
\ie
\Phi(t,r,\theta,\varphi) = \sum^{\infty}_{l=0}\sum^{l}_{m=-l}r^{-1}\Psi_{lm}(t,r)Y_{lm}(\theta,\varphi),\label{d1e2c3o4m5p6o7sition}
\fe
Here, the spherical harmonics are represented by \( Y_{lm}(\theta,\varphi) \). When we incorporate the decomposition of the scalar field, as outlined in Eq. (\ref{d1e2c3o4m5p6o7sition}), into Eq. (\ref{KleinGordon}), the equation assumes a Schrödinger--like form. This transformation endows the equation with wave--like characteristics, making it particularly well--suited for working on it
\ie
-\frac{\partial^{2} \Psi}{\partial t^{2}}+\frac{\partial^{2} \Psi}{\partial r^{*2}} + V_{eff}(r^{*})\Psi = 0.\label{s1c2h3o4r5din11ger}
\fe
The potential \(V_{\text{eff}}\) is often designated as the \textit{Regge--Wheeler} potential or the effective potential, capturing critical aspects of the black hole geometric structure. Additionally, we employ the tortoise coordinate \( r^{*} \), which spans the entire extent of spacetime and approaches \( r^{*} \rightarrow \pm \infty \). This coordinate is computed using the equation \(\mathrm{d} r^{*} = \sqrt{[1/f(r)^{2}]}\mathrm{d}r\), as elaborated below:
\ie
r^{*}= -2 (\lambda -1)^2 M+2 (\lambda -1)^2 M \ln [2 (\lambda -1) M+r] + r(1-\lambda).
\fe

\begin{figure}
    \centering
    \includegraphics[scale=0.35]{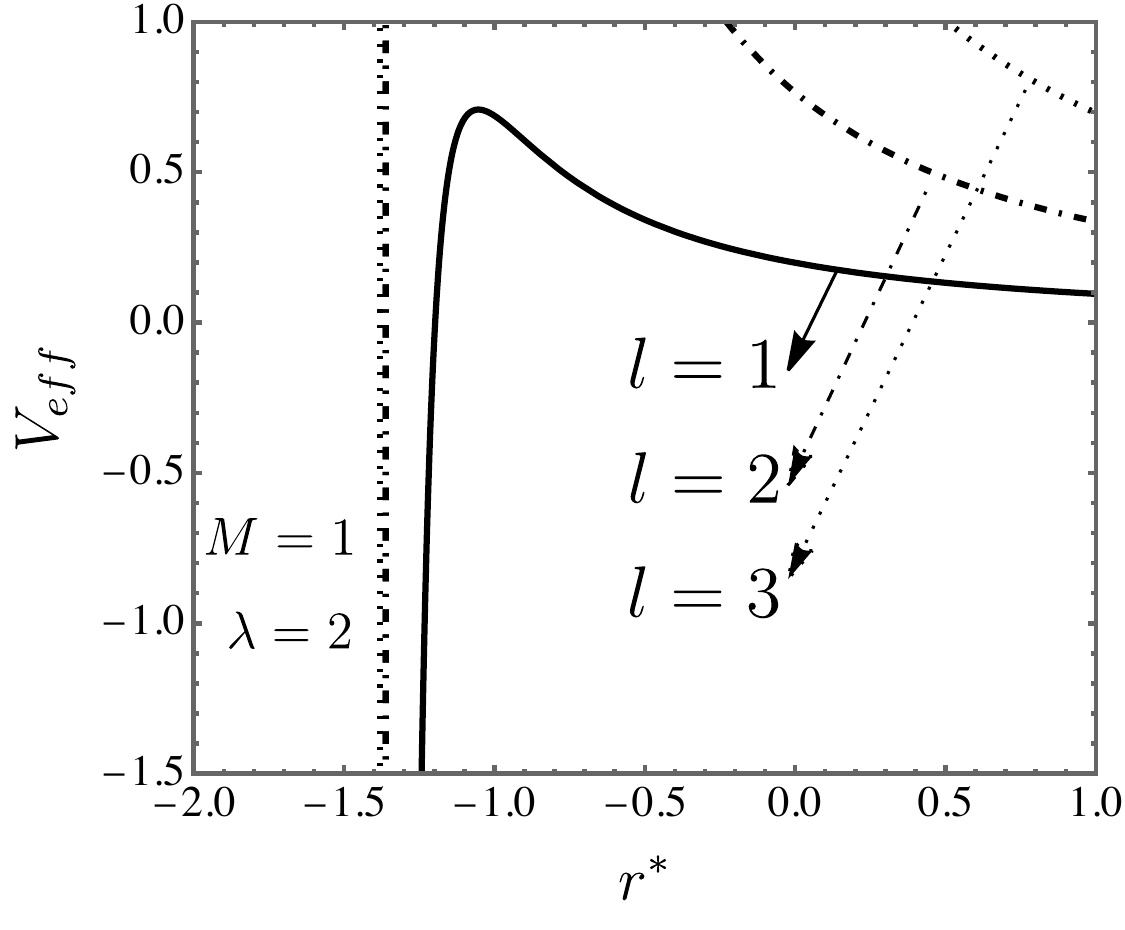}
    \caption{The effective potential $V_{eff}$ is depicted as a function of the tortoise coordinate $r^{*}$, considering different values of $l$.}
    \label{asdasd}
\end{figure}

Through algebraic manipulation, the effective potential can be reformulated as follows:
\ie
V_{eff} = f(r) \left(\frac{l (l+1)}{r^2}+\frac{2 M}{r^3}\right). \label{potenrtialeffecttta}
\fe
To facilitate a more nuanced understanding of Eq. \eqref{potenrtialeffecttta}, we direct attention to Fig. \ref{asdasd}. The figure reveals that a barrier--like configuration takes shape when, for example, \(M\) and \(\lambda=2\) are considered. Moreover, an increase in \( l \) is accompanied by a substantial elevation in the value of \( V_{eff} \).

Recent literature has seen similar examinations of quasinormal modes within various frameworks, including non--commutativity \cite{heidari2023gravitational,heidari2023exploring}, bumblebee gravity \cite{hassanabadi2023gravitational}, regular black holes \cite{aa2023analysis,aa2023implications}, and more. For the computation of these quasinormal modes, our methodology centers on the WKB approach. Our principal objective is to derive stationary solutions for the system under study. To realize this aim, we posit that the function \(\Psi(t,r)\) can be expressed as \(\Psi(t,r) = e^{-i\omega t} \psi(r)\), where \(\omega\) denotes the frequency. Employing this formulation allows us to readily isolate the time--independent component of Eq. (\ref{s1c2h3o4r5din11ger}) through the procedure detailed subsequently:
\ie
\frac{\partial^{2} \psi}{\partial r^{*2}} - \left[  \omega^{2} - V_{eff}(r^{*})\right]\psi = 0.
\fe


The WKB method, initially pioneered by Schutz and Will \cite{schutz1985black}, has become an invaluable tool for identifying quasinormal modes, particularly in the realm of particle scattering near black holes. This methodology has undergone refinements over time, most notably due to substantial contributions by Konoplya \cite{konoplya2003quasinormal, konoplya2004quasinormal}. It is important to acknowledge that the efficacy of this technique hinges on the potential adopting a barrier--like shape, which ultimately flattens to constant values as \( r^{*} \to \pm \infty \). By aligning the terms in the power of solution series with the peak potential turning points, quasinormal modes can be accurately computed. Given these conditions, the formula developed by Konoplya is articulated as follows:
\ie
\frac{i(\omega^{2}_{n}-V_{0})}{\sqrt{-2 V^{''}_{0}}} - \sum^{6}_{j=2} \Lambda_{j} = n + \frac{1}{2}.
\fe
In essence, Konoplya's framework for determining quasinormal modes comprises multiple elements. Specifically, the term \( V^{''}_{0} \) represents the second derivative of the potential, evaluated at its apex \( r_{0} \). Furthermore, the constants \( \Lambda_{j} \) are shaped by both the effective potential and its derivatives at this pinnacle. It is worth highlighting that recent advancements in the field have led to the development of a 13th--order WKB approximation, initiated by Matyjasek and Opala \cite{matyjasek2017quasinormal}. This progress significantly enhances the accuracy of quasinormal frequency calculations.

It is imperative to note that the quasinormal frequencies pertinent to the scalar field exhibit a negative imaginary component. This distinctive attribute signifies that these modes are subject to exponential decay over temporal intervals, thereby representing the mechanism of energy dissipation via scalar waves. This characteristic aligns coherently with extant scholarly literature that has investigated scalar, electromagnetic, and gravitational perturbations within spherically symmetric setup \cite{konoplya2011quasinormal,berti2009quasinormal,chen2023quasinormal}.

In Tables \ref{l00}, \ref{l11}, and \ref{l22}, we exhibit the behavior of \textit{quasinormal} modes, for different values of $l$, as they evolve with changes in two key parameters, namely, $M$ and $\lambda$, by employing the sixth--order WKB approximation. Remarkably, a consistent trend emerges across all cases: as $\lambda$ increases, these frequencies exhibit a notable increase in damping. In contrast, when we increment the mass parameter, we observe a contrasting behavior: the modes exhibit reduced damping.

\begin{table}[!h]
\begin{center}
\caption{\label{l00}Utilizing the sixth--order WKB approximation, we demonstrate the quasinormal frequencies corresponding to varying values of \( M \) and \( \lambda \), specifically for \( l=0 \).}
\begin{tabular}{c| c | c | c} 
 \hline\hline\hline 
 $M$ \,\,  $\lambda$  & $\omega_{0}$ & $\omega_{1}$ & $\omega_{2}$  \\ [0.2ex] 
 \hline 
 1.0, \, 0.00 & 0.110464 - 0.100819$i$ & 0.089023 - 0.344552$i$ & 0.191783 - 0.476507$i$ \\ 
 
 1.0, \, 0.10  & 0.136377 - 0.124466$i$  & 0.109907 - 0.425363$i$ & 0.236779 - 0.588257$i$  \\
 
 1.0, \, 0.11  & 0.139431 - 0.127304$i$ & 0.112340 - 0.435169$i$  &  0.241795 - 0.602365$i$  \\
 
 1.0, \, 0.12  & 0.142685 - 0.130153$i$  & 0.115034 - 0.444630$i$ &  0.248184 - 0.614008$i$  \\
 
 1.0, \, 0.13  & 0.145994 - 0.133153$i$ & 0.117709 - 0.454849$i$ & 0.254012 - 0.627978$i$ \\
 
 1.0, \, 0.14  & 0.149360 - 0.136312$i$ & 0.120373 - 0.465837$i$ & 0.259346 - 0.644182$i$  \\
 
 1.0, \, 0.15  & 0.152878 - 0.139554$i$ & 0.12319 - 0.476984$i$ & 0.265269 - 0.659956$i$ \\
 
 1.0, \, 0.16  & 0.156544 - 0.142893$i$ & 0.126151 - 0.488373$i$ & 0.271701 - 0.675582$i$  \\
 
 1.0, \, 0.17  & 0.160371 - 0.146327$i$ & 0.129268 - 0.499981$i$ & 0.278682 - 0.690960$i$  \\
  
 1.0, \, 0.18  & 0.164315 - 0.149909$i$ & 0.132448 - 0.512204$i$ & 0.285565 - 0.707774$i$ \\
 
  1.0, \, 0.19  & 0.168411 - 0.153622$i$ & 0.135773 - 0.524814$i$ & 0.292906 - 0.724792$i$  \\
  
 1.0, \, 0.20  & 0.172607 - 0.157523$i$ & 0.139113 - 0.538309$i$ & 0.299758 - 0.744307$i$ \\
   [0.2ex] 
 \hline \hline \hline 
  $M$ \,\,  $\lambda$  & $\omega_{0}$ & $\omega_{1}$ & $\omega_{2}$  \\ [0.2ex] 
 \hline 
 1.0, \, 0.1 & 0.1363770 - 0.1244660$i$ & 0.1099070 - 0.4253630$i$ & 0.2367790 - 0.5882570$i$ \\ 
 
 2.0, \, 0.1  & 0.0681880 - 0.0622330$i$ & 0.0549530 - 0.2126820$i$ & 0.1183900 - 0.2941280$i$ \\
 
 3.0, \, 0.1  & 0.0454482 - 0.0414980$i$ & 0.0366170 - 0.1418610$i$   &  0.0788040 - 0.1963900$i$  \\
 
 4.0, \, 0.1  & 0.0340940 - 0.0311160$i$ & 0.0274760 - 0.1063410$i$ &  0.0591948 - 0.1470640$i$  \\
 
 5.0, \, 0.1  & 0.0272707 - 0.0248976$i$ & 0.0219733 - 0.0851046$i$ & 0.0473030 - 0.1177820$i$  \\
 
 6.0, \, 0.1  & 0.0227320 - 0.0207421$i$ & 0.0183228 - 0.0708752$i$ & 0.0394972 - 0.0979586$i$  \\
 
 7.0, \, 0.1  & 0.0194746 - 0.0177881$i$ & 0.0156873 - 0.0608201$i$ & 0.0337353 - 0.0842628$i$ \\
 
 8.0, \, 0.1  & 0.0170472 - 0.0155583$i$  & 0.0137384 - 0.0531704$i$ & 0.0295974 - 0.0735321$i$ \\
 
 9.0, \, 0.1  & 0.0151562 - 0.0138266$i$ & 0.0122178 - 0.0472394$i$ & 0.0263493 - 0.0652601$i$  \\
  
 10.0, 0.1  & 0.0136372 - 0.0124471$i$ & 0.0109901 - 0.042539$i$  & 0.0236742 - 0.0588351$i$ \\
   [0.2ex] 
 \hline \hline \hline 
\end{tabular}
\end{center}
\end{table}

\begin{table}[!h]
\begin{center}
\caption{\label{l11}By applying the sixth--order WKB approximation, we elucidate the quasinormal frequencies associated with diverse values of \( M \) and \( \lambda \), with a specific focus on cases where \( l=1 \).}
\begin{tabular}{c| c | c | c} 
 \hline\hline\hline 
 $M$ \,\,  $\lambda$  & $\omega_{0}$ & $\omega_{1}$ & $\omega_{2}$  \\ [0.2ex] 
 \hline 
 1.0, \, 0.00 & 0.292910 - 0.097761$i$ & 0.264471 - 0.306518$i$  & 0.231014 - 0.542166$i$  \\ 
 
 1.0, \, 0.10  & 0.345670 - 0.120878$i$ & 0.309733 - 0.380212$i$ & 0.270155 - 0.673692$i$ \\
 
 1.0, \, 0.11  & 0.351810 - 0.123630$i$ & 0.314977 - 0.389000$i$  &  0.274720 - 0.689366$i$   \\
 
 1.0, \, 0.12  & 0.358132 - 0.126477$i$ & 0.320369 - 0.398100$i$ &  0.279412 - 0.705614$i$  \\
 
 1.0, \, 0.13  & 0.364646 - 0.129424$i$ & 0.325920 - 0.407519$i$ & 0.284254 - 0.722420$i$ \\
 
 1.0, \, 0.14  & 0.371359 - 0.132474$i$ & 0.331636 - 0.417272$i$ & 0.289249 - 0.739817$i$  \\
 
 1.0, \, 0.15  & 0.378280 - 0.135634$i$ & 0.337523 - 0.427378$i$ & 0.294400 - 0.757846$i$ \\
 
 1.0, \, 0.16  & 0.385418 - 0.138907$i$ & 0.343589 - 0.437854$i$ & 0.299715 - 0.776531$i$ \\
 
 1.0, \, 0.17  & 0.392782 - 0.142301$i$ & 0.349839 - 0.448720$i$ & 0.305199 - 0.795918$i$  \\
  
 1.0, \, 0.18  & 0.400383 - 0.145820$i$ & 0.356286 - 0.459989$i$ & 0.310869 - 0.816013$i$ \\
 
  1.0, \, 0.19  & 0.408231 - 0.149471$i$ & 0.362936 - 0.471686$i$ & 0.316729 - 0.836866$i$ \\
  
   1.0, \, 0.20  & 0.416338 - 0.153261$i$ & 0.369797 - 0.483834$i$ & 0.322785 - 0.858524$i$ \\
   [0.2ex] 
 \hline \hline \hline 
  $M$ \,\,  $\lambda$  & $\omega_{0}$ & $\omega_{1}$ & $\omega_{2}$  \\ [0.2ex] 
 \hline 
 1.0, \, 0.1 & 0.345670 - 0.120878$i$ & 0.309733 - 0.380212$i$ & 0.270155 - 0.673692$i$  \\ 
 
 2.0, \, 0.1  & 0.172835 - 0.060438$i$ & 0.154867 - 0.190106$i$ & 0.135077 - 0.336846$i$ \\
 
 3.0, \, 0.1  & 0.115223 - 0.040292$i$ & 0.103245 - 0.126737$i$  &  0.090052 - 0.224563$i$  \\
 
 4.0, \, 0.1  & 0.086417 - 0.030219$i$ & 0.077433 - 0.095053$i$ &  0.067538 - 0.168423$i$  \\
 
 5.0, \, 0.1  & 0.069134 - 0.024175$i$ & 0.061947 - 0.076042$i$  & 0.054031 - 0.134736$i$   \\
 
 6.0, \, 0.1  & 0.057611 - 0.020146$i$ & 0.051622 - 0.0633685$i$ & 0.045026 - 0.112281$i$  \\
 
 7.0, \, 0.1  & 0.049381 - 0.017268$i$ & 0.044247 - 0.054316$i$ & 0.038593 - 0.096241$i$ \\

 8.0, \, 0.1  & 0.043208 - 0.015109$i$ & 0.038716 - 0.047526$i$ & 0.033769 - 0.084211$i$ \\
 
 9.0, \, 0.1  & 0.038407 - 0.013430$i$ & 0.034414 - 0.042245$i$ & 0.030017 - 0.074854$i$  \\
  
 10.0, 0.1  & 0.034567 - 0.012087$i$ & 0.030973 - 0.038021$i$ & 0.027015 - 0.067368$i$  \\
   [0.2ex] 
 \hline \hline \hline 
\end{tabular}
\end{center}
\end{table}

\begin{table}[!h]
\begin{center}
\caption{\label{l22}Utilizing the sixth--order WKB approximation, we present the quasinormal frequencies corresponding to various values of \( M \) and $\Lambda$, particularly when \( l=2 \).}
\begin{tabular}{c| c | c | c} 
 \hline\hline\hline 
 $M$ \,\,  $\lambda$  & $\omega_{0}$ & $\omega_{1}$ & $\omega_{2}$  \\ [0.2ex] 
 \hline 
 1.0, \, 0.00 & 0.483642 - 0.096766$i$ & 0.463847 - 0.295627$i$ & 0.430386 - 0.508700$i$ \\ 
 
 1.0, \, 0.10  & 0.568021 - 0.119536$i$ & 0.542536 - 0.365844$i$ & 0.500283 - 0.631162$i$ \\
 
 1.0, \, 0.11  & 0.577801 - 0.122246$i$ & 0.551626 - 0.374211$i$  &  0.508323 - 0.645777$i$  \\
 
 1.0, \, 0.12  & 0.587865 - 0.125048$i$ & 0.560972 - 0.382868$i$  &  0.516583 - 0.660905$i$  \\
 
 1.0, \, 0.13  & 0.598224 - 0.127948$i$ & 0.570587 - 0.391828$i$ & 0.525071 - 0.676570$i$ \\
 
 1.0, \, 0.14  & 0.608890 - 0.130951$i$ & 0.580479 - 0.401107$i$ & 0.533799 - 0.692794$i$  \\
 
 1.0, \, 0.15  & 0.619877 - 0.134060$i$ & 0.590662 - 0.410718$i$ & 0.542775 - 0.709607$i$ \\
 
 1.0, \, 0.16  & 0.631199 - 0.137281$i$ & 0.601147 - 0.420679$i$ & 0.552009 - 0.727038$i$ \\
 
 1.0, \, 0.17  & 0.642869 - 0.140620$i$ & 0.611946 - 0.431007$i$ & 0.561511 - 0.745117$i$  \\
  
 1.0, \, 0.18  & 0.654903 - 0.144082$i$  & 0.623074 - 0.441719$i$ & 0.571295 - 0.763873$i$ \\
 
1.0, \, 0.19  & 0.667317 - 0.147673$i$ & 0.634544 - 0.452835$i$ & 0.581368 - 0.783347$i$ \\
  
   1.0, \, 0.20  & 0.680128 - 0.151401$i$ & 0.646371 - 0.464375$i$ & 0.591747 - 0.803569$i$ \\
   [0.2ex] 
 \hline \hline \hline 
  $M$ \,\,  $\lambda$  & $\omega_{0}$ & $\omega_{1}$ & $\omega_{2}$  \\ [0.2ex] 
 \hline 
 1.0, \, 0.1 & 0.568021 - 0.119536$i$ & 0.542536 - 0.365844$i$ & 0.500283 - 0.631162$i$ \\ 
 
 2.0, \, 0.1  & 0.284011 - 0.059768$i$  & 0.271268 - 0.182922$i$ & 0.250142 - 0.315581$i$ \\
 
 3.0, \, 0.1  & 0.189340 - 0.039845$i$ & 0.180845 - 0.121948$i$  &  0.166761 - 0.210387$i$  \\
 
 4.0, \, 0.1  & 0.142005 - 0.029884$i$ & 0.135634 - 0.091461$i$ &  0.125071 - 0.157790$i$  \\
 
 5.0, \, 0.1  & 0.113604 - 0.023907$i$ & 0.108507 - 0.073168$i$ & 0.100057 - 0.126232$i$  \\
 
 6.0, \, 0.1  & 0.094670 - 0.019922$i$ & 0.090422 - 0.060974$i$ & 0.083380 - 0.105194$i$  \\
 
 7.0, \, 0.1  & 0.081145 - 0.017076$i$ & 0.077505 - 0.052263$i$ & 0.071469 - 0.090165$i$ \\
 
 8.0, \, 0.1  & 0.071002 - 0.014942$i$ & 0.067817 - 0.045730$i$ & 0.062535 - 0.078895$i$ \\
 
 9.0, \, 0.1  & 0.063113 - 0.013281$i$ & 0.060281 - 0.040649$i$ & 0.055587 - 0.070129$i$  \\
  
 10.0, 0.1  & 0.056802 - 0.011953$i$ & 0.054253 - 0.036584$i$ & 0.050028 - 0.063116$i$ \\
   [0.2ex] 
 \hline \hline \hline 
\end{tabular}
\end{center}
\end{table}



\subsection{With presence of cosmological constant}

In the presence of a cosmological constant, it becomes evident that the initial assumption of \(V' = 0\) does not yield a self--consistent solution that complies with all the equations of motion. As a result, we adopt the methodology described in Ref. \cite{maluf2021black} to relax the vacuum conditions to \(V = 0\) while allowing \(V' \neq 0\). The linear potential, \(V = \sigma X\), is often the most frequently discussed form that meets this criteria, where \(\sigma\) serves as a Lagrange multiplier field \cite{bluhm2008spontaneous}.

Taking the derivative of this potential with respect to \(X\) yields \(V'(X) = \sigma\). The equation of motion for the Lagrange multiplier \(\sigma\) constrains the theory to the potential extrema, which correspond to \(X = 0\). In this case, \(b_{\mu \nu}\) becomes the VEV of the KR field for the on--shell \(\sigma\).

It should be noted that, to maintain the positivity of the potential \(V\), the off--shell value of the Lagrange multiplier \(\sigma\) must share the same sign as \(X\). Although it is theoretically possible to expand the Lagrange multiplier field \(\sigma\) around its vacuum value, i.e., expressed as \(\sigma = \langle \sigma \rangle + \tilde{\sigma}\), and allow \(\langle \sigma \rangle\) to vary with spacetime coordinates, for the sake of simplicity we opt to set \(\tilde{\sigma} = 0\) through appropriate initial conditions. We further assume that \(\langle \sigma \rangle\) is a real constant. This ensures that the on--shell value of \(\sigma\), denoted as \(\sigma \equiv \langle \sigma \rangle\), is uniquely determined by the field equations \cite{bluhm2008spontaneous}.

Furthermore, after some algebraic manipulations, we get \cite{yang2023static}
\ie
A(r)=\frac{1}{B(r)}=\frac{1}{1-\lambda} - \frac{2M}{r} - \frac{\Lambda r^{2}}{3(1-\lambda)},\label{LapseF}
\fe
which yields
\ie
\label{cosmologicalconstant}
\mathrm{d}s^{2} = - \left( \frac{1}{1-\lambda} - \frac{2M}{r} - \frac{\Lambda r^{2}}{3(1-\lambda)}   \right) \mathrm{d}t^{2} + \frac{\mathrm{d}r^{2}}{\frac{1}{1-\lambda} - \frac{2M}{r} - \frac{\Lambda r^{2}}{3(1-\lambda)} } + r^{2}\mathrm{d}\theta^{2} + r^{2} \sin^{2}\mathrm{d}\varphi^{2}.
\fe
Above expression gives rise to the following physical horizon
\ie
\begin{split}
r_{\Lambda \neq 0} = \frac{\Lambda +\left(\sqrt{\Lambda ^3 \left(9 (\lambda -1)^2 \Lambda  M^2-1\right)}+3 (\lambda -1) \Lambda ^2 M\right)^{2/3}}{\Lambda  \sqrt[3]{\sqrt{\Lambda ^3 \left(9 (\lambda -1)^2 \Lambda  M^2-1\right)}+3 (\lambda -1) \Lambda ^2 M}}.
\end{split}
\fe


\subsubsection{Photon sphere and shadows}

\begin{figure}
    \centering
    \includegraphics[scale=0.24]{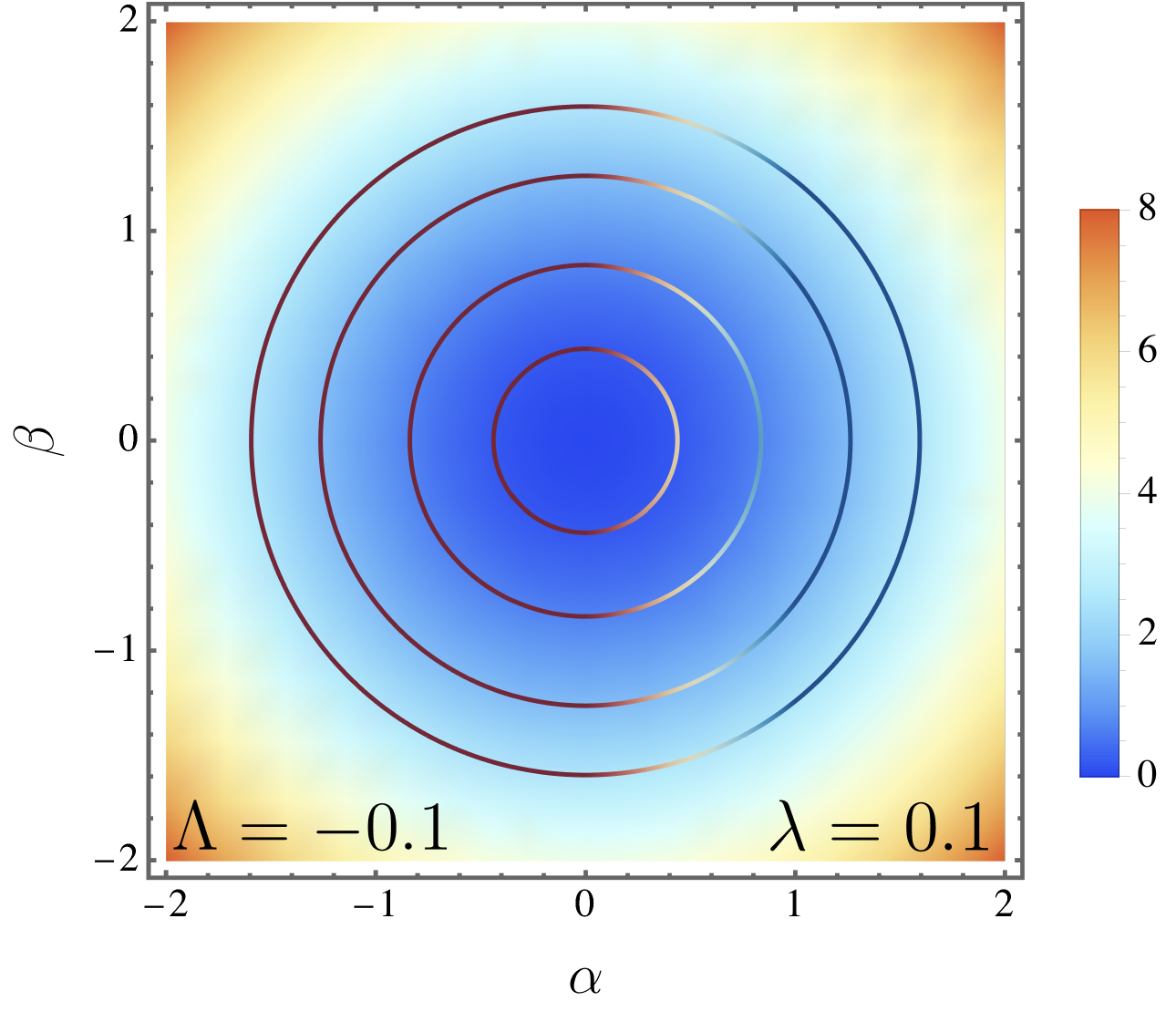}
    \includegraphics[scale=0.24]{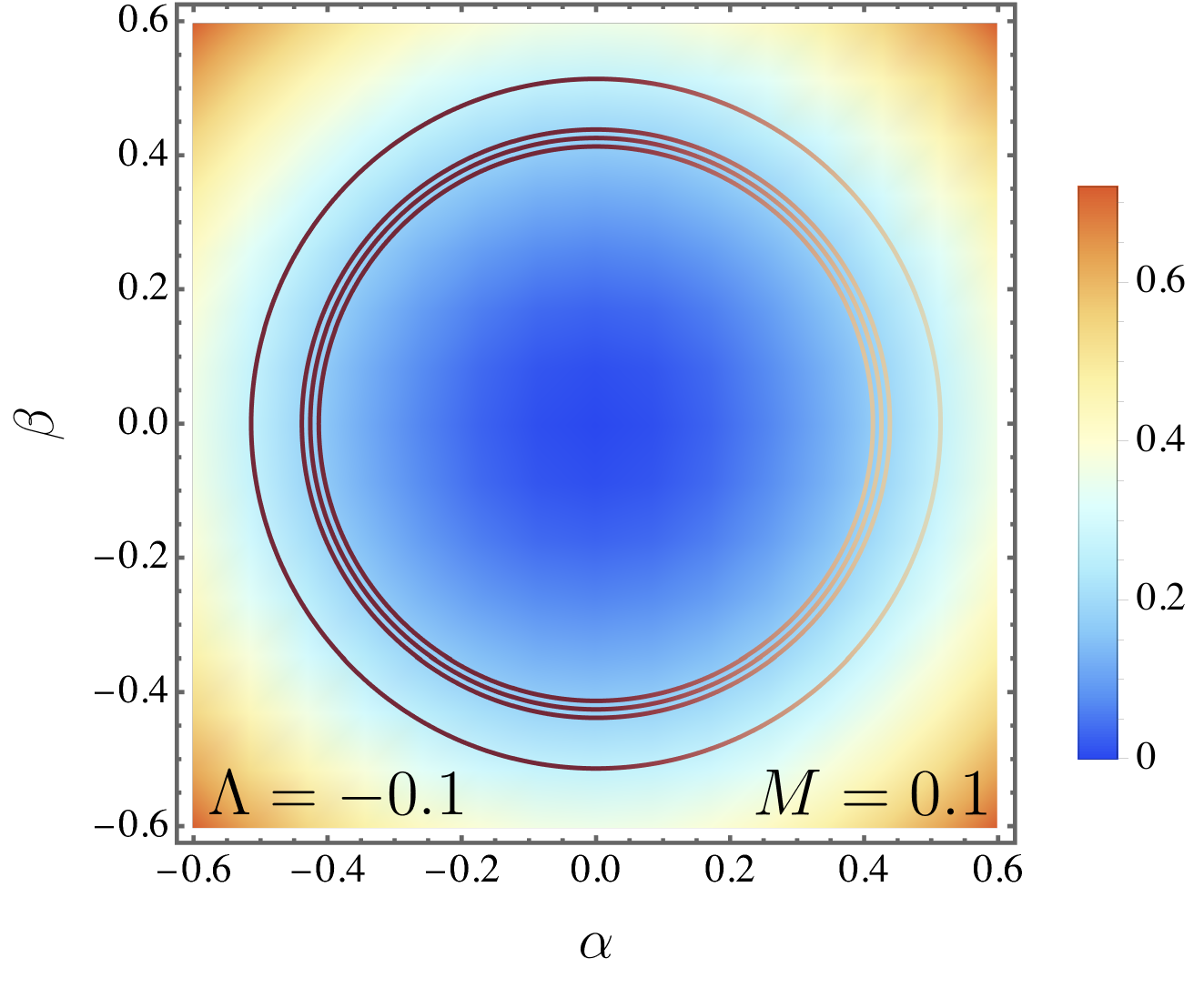}
    \includegraphics[scale=0.24]{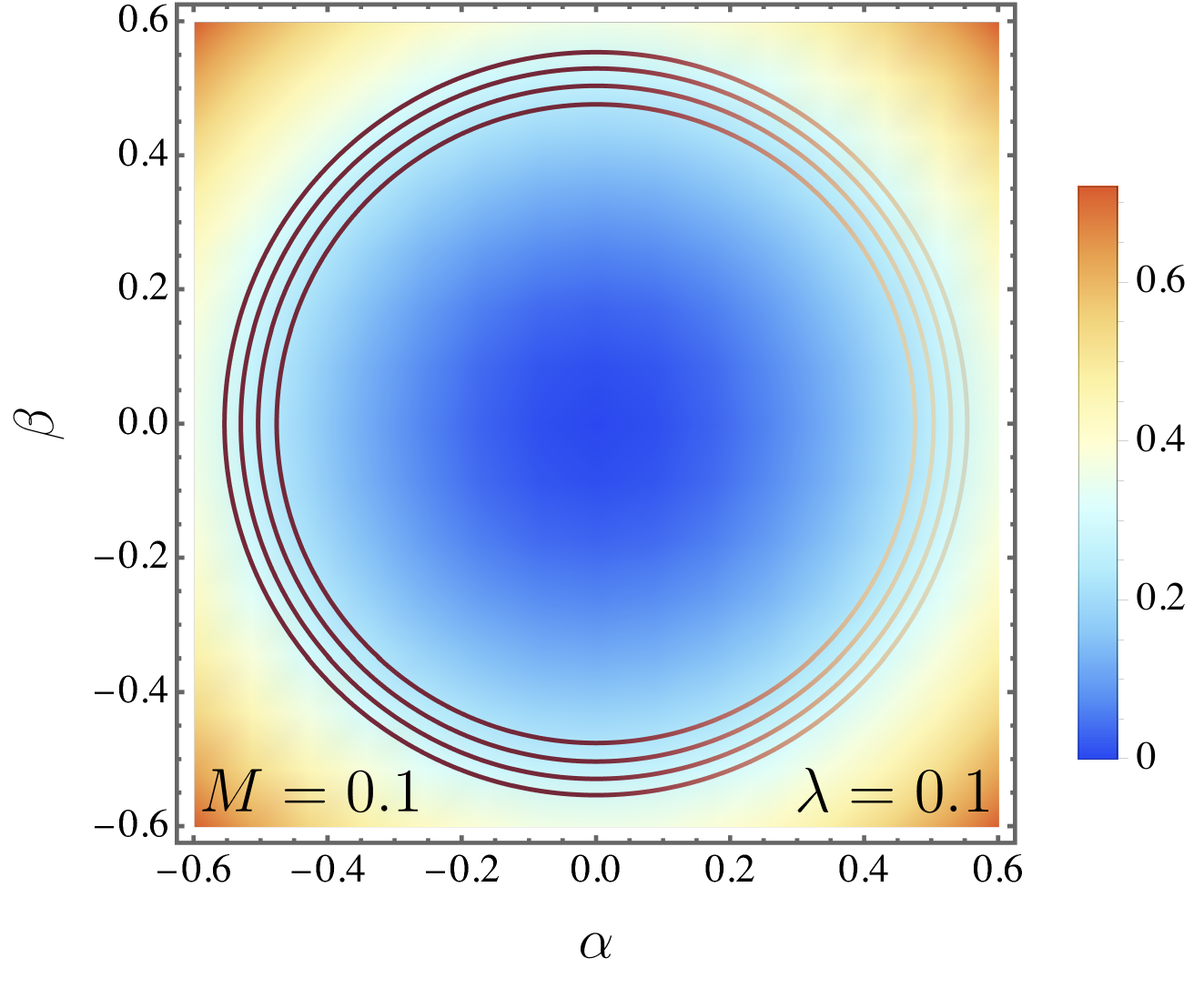}
    \caption{The diagram showcases an array of circles, each representing the shadow formations resulting from variations in the parameters \( M \), \( \lambda \), and \( \Lambda \). In the upper--left quadrant, these circles correspond to shadows generated by \( M \) values ranging from $0.1$ (innermost circle) to $0.4$ (outermost circle), increasing incrementally by integers. In this context, \( \lambda \) remains fixed at $0.1$, while \( \Lambda \) is set at $-0.1$. Conversely, in the upper-right quadrant, the circles portray shadows influenced by varying \( \lambda \) values. These \( \lambda \) values range from $0.1$ (innermost circle) to $0.175$ (outermost circle), with \( M \) maintained at $0.1$ and \( \Lambda \) at $-0.1$. In this arrangement, each successive circle exhibits a \( \lambda \) value incremented by $0.025$, and the shadow radius diminishes as \( \lambda \) increases. In the lower section of the illustration, the circles are determined by a decrement of \( -0.1 \) in \( \Lambda \) for each subsequent representation. The span commences with \( \Lambda = -0.1 \) for the innermost circle and descends to \( \Lambda = -0.4 \) for the outermost circle. }
    \label{ssssshadows23c45omplete}
\end{figure}

Remarkably, it is important to mention that the cosmological constant does not affect the photon sphere radius. The same facet, in the other hand, can not be inferred to the shadow radius. In addition, in the case of substantial cosmic distances and the presence of a cosmological constant, it has been demonstrated that the size of a black hole's shadow can explicitly vary with respect to the radial coordinate of a distant observer denoted as $r_O$. This relationship is detailed in \cite{Gonzalez:2023rsd} and can be represented as:
\begin{eqnarray}
\mathcal{R}_{\Lambda\neq 0} = r_{c} \sqrt{\frac{f(r_O)}{f(r_{c})}} = \frac{3 (1-\lambda) M \sqrt{\frac{-6 \lambda  M+6 M+\Lambda  r_{O}^3-3 r_{O}}{(\lambda -1) r_{O}}}}{\sqrt{\frac{1}{1-\lambda }+9 (\lambda -1) \Lambda  M^2}}\,.
\label{eq:rshnotasymptoticallyflat}
\end{eqnarray}
In Fig. \ref{ssssshadows23c45omplete}, it is exhibited the shadows for different values of varying the parameters \( M \), \( \lambda \), and $\Lambda$. In the left section of the diagram, shadows (circles) are formed by varying the mass parameter \( M \), i.e., $0.1, 0.2, 0.3$ and $0.4$. Here, the innermost circle corresponds to \( M = 0.1 \) and the outermost to \( M = 0.4 \). For these circles, \( \lambda \) remains constant at $0.1$, while \( \Lambda \) is set to $-0.1$. The increment of $M$ is responsible for making larger values of shadow radius.

In contrast, the middle section features circles whose properties are dictated by changes in the coupling constant \( \lambda \). In this setting, \( \lambda \) varies from $0.125$ (outermost circle) to $0.2$ (outermost circle), with each consecutive circle incremented by $0.025$ in \( \lambda \). Notably, the radius of the shadow shrinks as \( \lambda \) increases, while \( M \) and \( \Lambda \) are fixed at $0.1$ and $-0.1$, respectively. Finally, the right section of the diagram, the circles are characterized by decrements of $-0.1$ in \( \Lambda \) for each subsequent plot. The value of \( \Lambda \) starts at $-0.1$ for the innermost circle and decreases to $-0.4$ for the outermost circle.

\subsubsection{Constraints on the parameters of antisymmetric black hole with the EHT observations Sgr A*}

In this section, we derive constraints on the parameters of antisymmetric black holes by analyzing observational data obtained from the Event Horizon Telescope (EHT) \cite{118,119,120,121} focusing on Sagittarius $A^{*}$ shadow images. The shadow image's radius variation for Sgr. $A^{*}$ is quantified with uncertainties at $1\sigma$ and $2\sigma$ levels.

\begin{figure}[htbp]
		\centerline{
			\includegraphics[scale = 0.46]{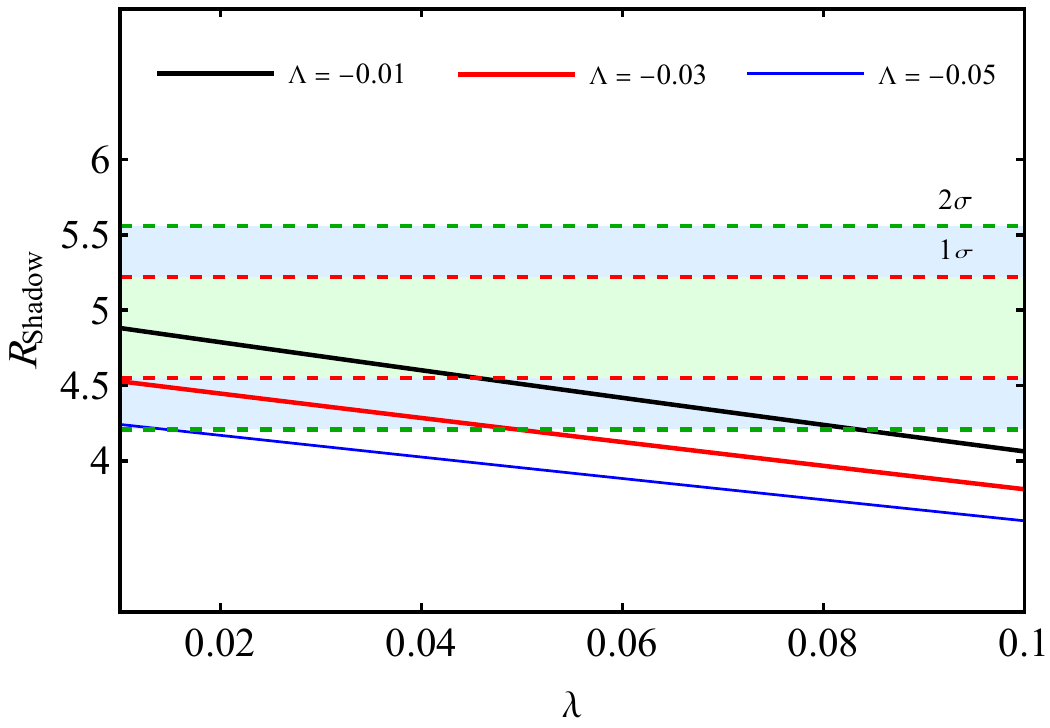}\hspace{0.5cm}
			\includegraphics[scale = 0.46]{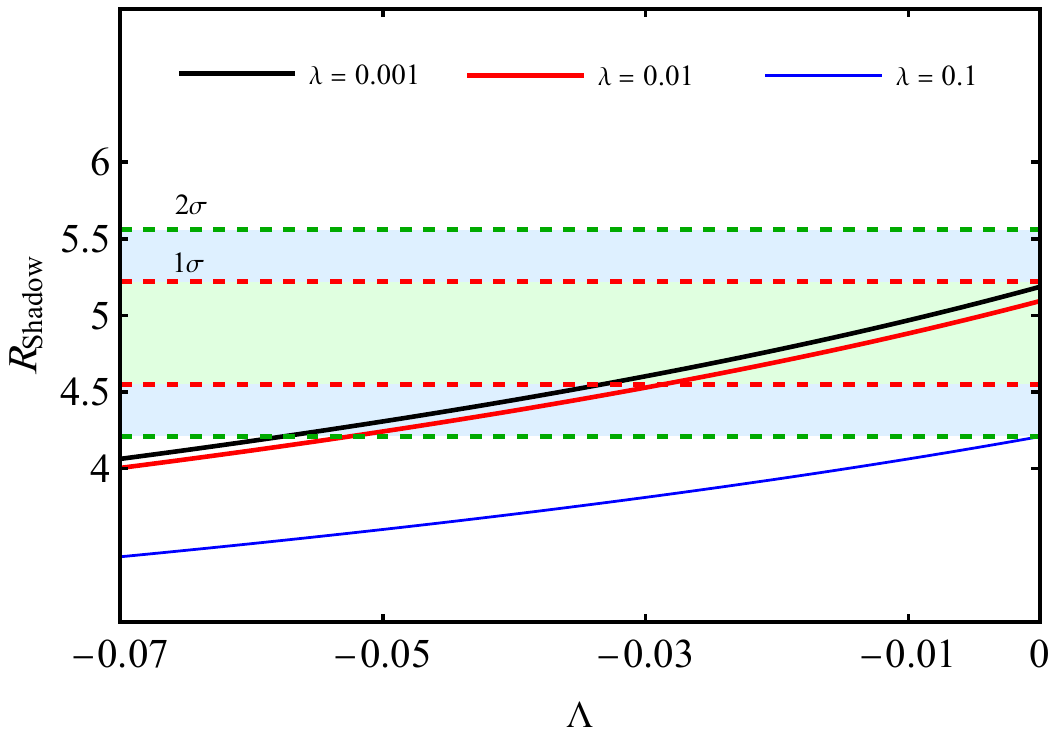}} \vspace{-0.2cm}
		\caption{The shadow radius plots according to $\lambda$ are shown in different color lines for $\Lambda=-0.01, -0.03, -0.05$ in left panel. On the right panel, color lines represent the shadow radius versus $\Lambda$ for $\lambda=0.001, 0.01, 0.1$. In both panels two pairs of dashed lines for experimental constraints are shown.}
		\label{constraints}
	\end{figure}

Fig. \ref{constraints} illustrates the relationship between the shadow and the parameters $\Lambda$ and $\lambda$. Tables \ref{cons1} and \ref{cons2} present the acceptable parameter constraints for $\lambda$ and $\Lambda$ based on experimental limits at $1\sigma$ and $2\sigma$ levels, respectively.

\begin{table}[!ht]
	\centering
	\caption{\label{cons1}The results of the constraints for $\lambda$ for some values of $\Lambda$ based on $\sigma1$ and $\sigma2$.}
	\begin{tabular}{|l|l|l|l|l|}
		\hline
		$\Lambda$ & ~ & -0.01 & -0.03 & -0.05 \\ \hline
		$\sigma1$ & Upper band for 	$\lambda$ & 0.08 & 0.049 & 0.014 \\ \hline
		~ & Lower band for 	$\lambda$ & - & - & - \\ \hline
		$\sigma$2 & Upper band for $\lambda$& 0.045 & 0.007 & - \\ \hline
		~ & Lower band for $\lambda$ & - & - & - \\ \hline
	\end{tabular}
	
\end{table}

\begin{table}[!ht]
	\centering
	\caption{\label{cons2}The results of the constraints for $\Lambda$ for some values of $\lambda$ based on $\sigma1$ and $\sigma2$.}
	\begin{tabular}{|l|l|l|l|l|}
		\hline
		$\lambda$ & ~ & 0.001 & 0.01 & 0.1 \\ \hline
		$\sigma1$ & Upper band for $\Lambda$ & -0.057 & -0.052 & - \\ \hline
		~ & Lower band for $\Lambda$ & - & - & - \\ \hline
		$\sigma2$ & Upper band for $\Lambda$ & -0.033 & -0.028 & - \\ \hline
		~ & Lower band for $\Lambda$ & - & - & - \\ \hline
	\end{tabular}
\end{table}

\subsubsection{Quasinormal modes}

From Eq. (\ref{cosmologicalconstant}), 
the \textit{Regge--Wheeler} potential reads
\ie
V_{eff(\Lambda)}= f(r)\left(\frac{\frac{2 M}{r^2}-\frac{2 \Lambda  r}{3 (1-\lambda)}}{r}+\frac{l (l+1)}{r^2}\right)\label{effpot}.
\fe
Analogously what we have accomplished in the previous sections, here we present the analysis of the \textit{quasinormal} modes, considering $\Lambda \neq 0$ instead. The main features ascribed to this, will be shown as follows.

\begin{figure}
    \centering
    \includegraphics[scale=0.4]{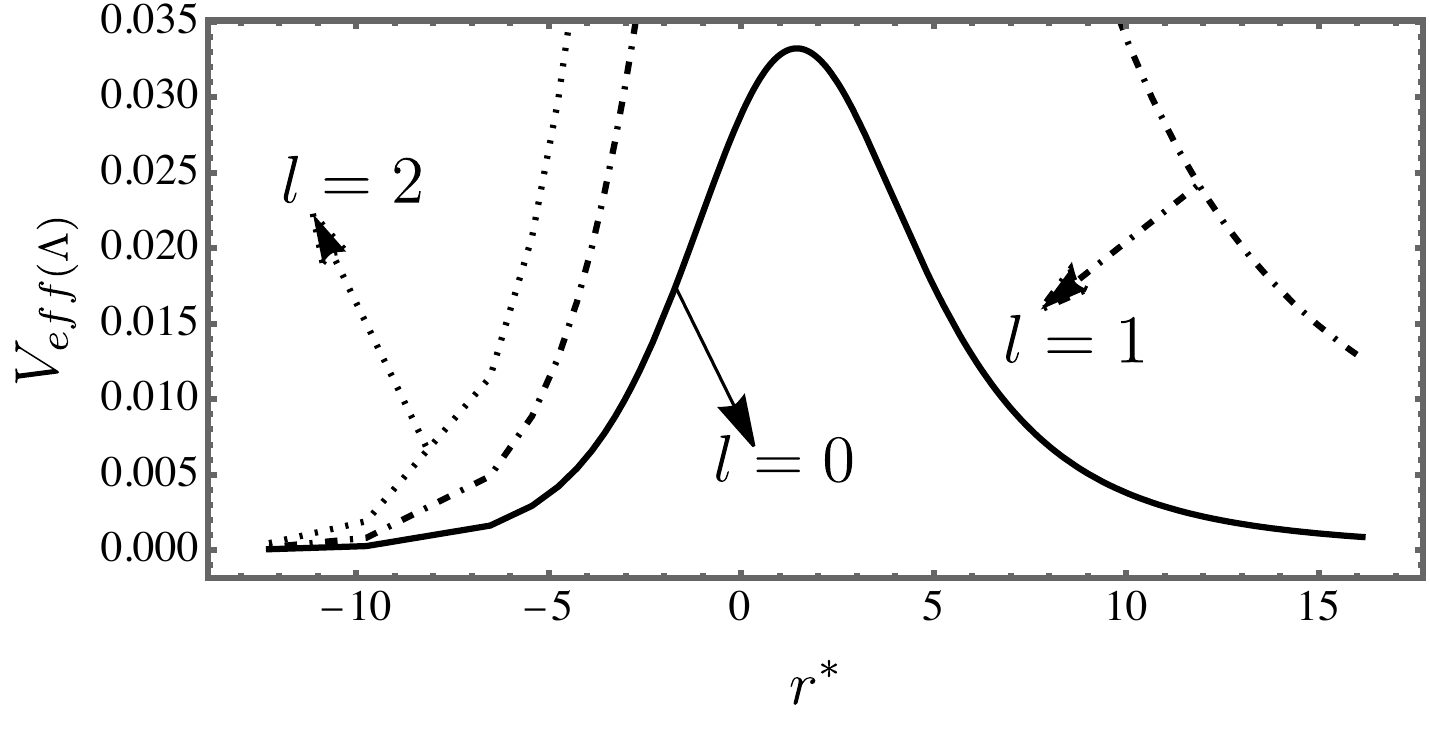}
    \caption{The effective potential $V_{eff(\Lambda)}$ is depicted as a function of the tortoise coordinate $r^{*}$, considering different values of $l$ and fixed values of $\Lambda=-0.01$, $\lambda = 0.1$, and $M=1$.}
    \label{potentiallastoasdne}
\end{figure}

In Fig. \ref{potentiallastoasdne}, we observe
the effective potential, denoted as $V_{eff(\Lambda)}$, plotted against the tortoise coordinate $r^{*}$. This representation explores various values of $l$ while maintaining constant values for the parameters $\Lambda=-0.01$, $\lambda=0.1$, and $M=1$.

In Tables \ref{l0}, \ref{l1}, and \ref{l2}, we present the \textit{quasinormal} frequencies computed for various values of the quantum number $l$, utilizing the sixth--order WKB method. Additionally, we conduct a comprehensive analysis of the behavior of these modes as they depend on the parameters $M$, $\lambda$, and $\Lambda$. It is noteworthy that across all considered values of $l$, i.e., specifically, $l=0$, $l=1$, and $l=2$, the observed trends in the modes, with respect to variations in $M$ and $\lambda$, closely resemble those identified in the previous case, as one might logically anticipate. It is important to highlight that as the cosmological constant $\Lambda$ attains increasingly negative values, the quasinormal frequencies demonstrate a damper aspect.

It is worth noting that a remarkable feature emerges when $l=0$: certain modes exhibit instability for the considered parameter values. However, as $l$ varies, these instabilities are no longer maintained.

\begin{table}[!h]
\begin{center}
\caption{\label{l0}Utilizing the sixth--order WKB approximation, we demonstrate the quasinormal frequencies corresponding to varying values of \( M \), \( \lambda \), and $\Lambda$ specifically for \( l=0 \).}
\begin{tabular}{c| c | c | c} 
 \hline\hline\hline 
 \!\!\!\! $M$ \,\,\,\,\,\,  $\lambda$  \,\,\,\, \,\, $\Lambda$  & $\omega_{0}$ & $\omega_{1}$ & $\omega_{2}$  \\ [0.2ex] 
 \hline 
 1.00, \, 0.10, \, -0.01 & 0.140914 - 0.124590$i$ & 0.0356274 - 0.589974$i$ & \text{Unstable} \\ 
 
 1.01, \, 0.10, \, -0.01 & 0.139526 - 0.123374$i$ & 0.0321124 - 0.593042$i$ & \text{Unstable}  \\
 
 1.02, \, 0.10, \, -0.01 & 0.138166 - 0.122178$i$ & 0.0285944 - 0.596445$i$  &  \text{Unstable}  \\
 
 1.03, \, 0.10, \, -0.01 & 0.136818 - 0.121013$i$  & 0.0250711 - 0.600259$i$ &  \text{Unstable}  \\
 
 1.04, \, 0.10, \, -0.01 & 0.135483 - 0.119876$i$ & 0.0215458 - 0.604477$i$ & \text{Unstable} \\
 
 1.05, \, 0.10, \, -0.01  & 0.134172 - 0.118756$i$ & 0.0180223 - 0.609043$i$ & \text{Unstable}  \\
 
 1.06, \, 0.10, \, -0.01 & 0.132868 - 0.117668$i$ & 0.0144995 - 0.614046$i$ & \text{Unstable} \\
 
 1.07, \, 0.10, \, -0.01 & 0.131578 - 0.116601$i$ & 0.0109807 - 0.619443$i$ & \text{Unstable}  \\
 
 1.08, \, 0.10, \, -0.01  & 0.130301 - 0.115558$i$ & 0.0074673 - 0.625253$i$ & \text{Unstable} \\
  
 1.09, \, 0.10, \, -0.01 & 0.129026 - 0.114544$i$ & 0.0039610 - 0.631505$i$  & \text{Unstable} \\

  1.10, \, 0.10, \, -0.01 & 0.127765 - 0.113549$i$ & 0.0004630 - 0.638159$i$ & \text{Unstable} \\
   [0.2ex] 
 \hline \hline \hline 
  \!\!\!\! $M$ \,\,\,\,\,\,  $\lambda$  \,\,\,\, \,\, $\Lambda$   & $\omega_{0}$ & $\omega_{1}$ & $\omega_{2}$  \\ [0.2ex] 
 \hline 
 1.00, \, 0.10,  \, -0.01 & 0.140914 - 0.124590$i$ & 0.0356274 - 0.589974$i$ & \text{Unstable} \\ 
 
 1.00, \, 0.11, \, -0.01 & 0.144078 - 0.127389$i$ & 0.0399678 - 0.593609$i$ & \text{Unstable} \\
 
 1.00, \, 0.12, \, -0.01 & 0.147355 - 0.130275$i$ & 0.0444021 - 0.597706$i$  &  \text{Unstable}  \\
 
 1.00, \, 0.13, \, -0.01 & 0.150718 - 0.133280$i$  & 0.0489170 - 0.602466$i$ & \text{Unstable}  \\
 
 1.00, \, 0.14, \, -0.01 & 0.154207 - 0.136370$i$ & 0.0535301 - 0.607694$i$ & \text{Unstable}  \\
 
 1.00, \, 0.15, \, -0.01 & 0.157812 - 0.139589$i$ & 0.0582345 - 0.613502$i$  & \text{Unstable}  \\
 
 1.00, \, 0.16, \, -0.01 & 0.161528 - 0.142928$i$ & 0.0630258 - 0.619972$i$ & \text{Unstable} \\
 
 1.00, \, 0.17, \, -0.01 & 0.165388 - 0.146375$i$ & 0.0679225 - 0.626949$i$ & \text{Unstable} \\
 
 1.00, \, 0.18, \, -0.01 & 0.169366 - 0.149968$i$ & 0.0729055 - 0.634656$i$ & \text{Unstable}  \\
  
 1.00, \, 0.19, \, -0.01 & 0.173483 - 0.153690$i$ & 0.0779843 - 0.643031$i$ & \text{Unstable}  \\

  1.00, \, 0.20, \, -0.01 & 0.177773 - 0.157550$i$ & 0.0831850 - 0.651918$i$ & \text{Unstable}  \\
   [0.2ex] 
 \hline \hline \hline 
  \!\!\!\! $M$ \,\,\,\,\,\,  $\lambda$  \,\,\,\, \,\, $\Lambda$  & $\omega_{0}$ & $\omega_{1}$ & $\omega_{2}$  \\ [0.2ex] 
 \hline 
 1.0, \, 0.1 \, -0.0010 & 0.136954 - 0.124470$i$ & 0.109389 - 0.428021$i$ & 0.222318 - 0.608012$i$ \\ 
 
 1.0, \, 0.1 \, -0.0011  & 0.137019 - 0.124463$i$ & 0.109298 - 0.428297$i$ & 0.220240 - 0.610849$i$ \\
 
 1.0, \, 0.1 \, -0.0012 & 0.137059 - 0.124479$i$ & 0.109150 - 0.428766$i$  &  0.217755 - 0.614682$i$  \\
 
 1.0, \, 0.1 \, -0.0013 & 0.137138 - 0.124459$i$ & 0.109060 - 0.428976$i$ & 0.215557 - 0.617572$i$  \\
 
 1.0, \, 0.1 \, -0.0014 & 0.137192 - 0.124463$i$ & 0.108914 - 0.429380$i$ & 0.212961 - 0.621467$i$  \\
 
 1.0, \, 0.1 \, -0.0015 & 0.137243 - 0.124469$i$ & 0.108751 - 0.429820$i$ & 0.210211 - 0.625691$i$  \\
 
 1.0, \, 0.1 \, -0.0016 & 0.137293 - 0.124475$i$ & 0.108577 - 0.430276$i$ & 0.207349 - 0.630143$i$ \\
 
 1.0, \, 0.1 \, -0.0017 & 0.137359 - 0.124467$i$ & 0.108418 - 0.430641$i$ & 0.204510 - 0.634402$i$ \\
 
 1.0, \, 0.1 \, -0.0018 & 0.137422 - 0.124462$i$ & 0.108242 - 0.431039$i$ & 0.201534 - 0.638973$i$  \\
  
 1.0, \, 0.1 \, -0.0019  & 0.137471 - 0.124470$i$ & 0.108027 - 0.431560$i$  & 0.198299 - 0.644263$i$ \\

 1.0, \, 0.1 \, -0.0020  & 0.137537 - 0.124461$i$ & 0.107832 - 0.431968$i$  & 0.195135 - 0.649239$i$ \\
   [0.2ex] 
 \hline \hline \hline 
\end{tabular}
\end{center}
\end{table}

\begin{table}[!h]
\begin{center}
\caption{\label{l1}Utilizing the sixth--order WKB approximation, we demonstrate the quasinormal frequencies corresponding to varying values of \( M \), \( \lambda \), and $\Lambda$ specifically for \( l=1 \).}
\begin{tabular}{c| c | c | c} 
 \hline\hline\hline 
 \!\!\!\! $M$ \,\,\,\,\,\,  $\lambda$  \,\,\,\, \,\, $\Lambda$  & $\omega_{0}$ & $\omega_{1}$ & $\omega_{2}$  \\ [0.2ex] 
 \hline 
 1.00, \, 0.10, \, -0.01 & 0.360556 - 0.123469$i$ & 0.317340 - 0.394362$i$ & 0.260939 - 0.717176$i$ \\ 
 
 1.01, \, 0.10, \, -0.01 & 0.357277 - 0.122294$i$ & 0.314274 - 0.390758$i$ & 0.257535 - 0.711608$i$  \\
 
 1.02, \, 0.10, \, -0.01 & 0.354065 - 0.121143$i$ & 0.311264 - 0.387230$i$  &  0.254151 - 0.706211$i$  \\
 
 1.03, \, 0.10, \, -0.01 & 0.350918 - 0.120015$i$ & 0.308308 - 0.383775$i$ & 0.250782 - 0.700980$i$  \\
 
 1.04, \, 0.10, \, -0.01 & 0.347835 - 0.118909$i$ & 0.305404 - 0.380392$i$ & 0.247428 - 0.695930$i$ \\
 
 1.05, \, 0.10, \, -0.01 & 0.344812 - 0.117824$i$ & 0.302549 - 0.377078$i$ & 0.244084 - 0.691043$i$  \\
 
 1.06, \, 0.10, \, -0.01 & 0.341849 - 0.116760$i$ & 0.299743 - 0.373832$i$ & 0.240748 - 0.686324$i$ \\
 
 1.07, \, 0.10, \, -0.01 & 0.338944 - 0.115716$i$ & 0.296984 - 0.370652$i$ &  0.237419 - 0.681770$i$  \\
 
 1.08, \, 0.10, \, -0.01  & 0.336095 - 0.114692$i$ & 0.294269 - 0.367536$i$ & 0.234093 - 0.677383$i$  \\
  
 1.09, \, 0.10, \, -0.01 & 0.333301 - 0.113687$i$ & 0.291597 - 0.364483$i$ & 0.230767 - 0.673162$i$ \\

  1.10, \, 0.10, \, -0.01 & 0.330560 - 0.112700$i$ & 0.288966 - 0.361491$i$ & 0.227439 - 0.669107$i$ \\
   [0.2ex] 
 \hline \hline \hline 
  \!\!\!\! $M$ \,\,\,\,\,\,  $\lambda$  \,\,\,\, \,\, $\Lambda$   & $\omega_{0}$ & $\omega_{1}$ & $\omega_{2}$  \\ [0.2ex] 
 \hline 
 1.00, \, 0.10,  \, -0.01 & 0.360556 - 0.123469$i$ & 0.317340 - 0.394362$i$ & 0.260939 - 0.717176$i$ \\ 
 
 1.00, \, 0.11, \, -0.01 & 0.366647 - 0.126212$i$ & 0.322546 - 0.403170$i$ & 0.265762 - 0.732657$i$ \\
 
 1.00, \, 0.12, \, -0.01 & 0.372920 - 0.129050$i$ & 0.327902 - 0.412285$i$  &  0.270718 - 0.748687$i$  \\
 
 1.00, \, 0.13, \, -0.01 & 0.379385 - 0.131988$i$ & 0.333415 - 0.421722$i$ & 0.275813 - 0.765296$i$  \\
 
 1.00, \, 0.14, \, -0.01 & 0.386048 - 0.135029$i$ & 0.339091 - 0.431496$i$ & 0.281054 - 0.782508$i$  \\
 
 1.00, \, 0.15, \, -0.01 & 0.392919 - 0.138178$i$ & 0.344938 - 0.441622$i$ & 0.286449 - 0.800349$i$  \\
 
 1.00, \, 0.16, \, -0.01 & 0.400006 - 0.141442$i$ & 0.350962 - 0.452119$i$ & 0.292004 - 0.818853$i$ \\
 
 1.00, \, 0.17, \, -0.01 & 0.407319 - 0.144826$i$ & 0.357172 - 0.463003$i$ & 0.297727 - 0.838048$i$ \\
 
 1.00, \, 0.18, \, -0.01 & 0.414869 - 0.148334$i$ & 0.363575 - 0.474295$i$ & 0.303627 - 0.857972$i$  \\
  
 1.00, \, 0.19, \, -0.01 & 0.422665 - 0.151975$i$ & 0.370180 - 0.486014$i$ & 0.309713 - 0.878656$i$  \\

  1.00, \, 0.20, \, -0.01 & 0.430720 - 0.155753$i$ & 0.376997 - 0.498183$i$ & 0.315994 - 0.900142$i$  \\
   [0.2ex] 
 \hline \hline \hline 
  \!\!\!\! $M$ \,\,\,\,\,\,  $\lambda$  \,\,\,\, \,\, $\Lambda$  & $\omega_{0}$ & $\omega_{1}$ & $\omega_{2}$  \\ [0.2ex] 
 \hline 
 1.0, \, 0.1 \, -0.0010 & 0.347181 - 0.121152$i$ & 0.310679 - 0.381601$i$ & 0.270925 - 0.676553$i$ \\ 
 
 1.0, \, 0.1 \, -0.0011  & 0.347332 - 0.121179$i$ & 0.310772 - 0.381739$i$ & 0.270992 - 0.676845$i$ \\
 
 1.0, \, 0.1 \, -0.0012 & 0.347482 - 0.121206$i$ & 0.310866 - 0.381878$i$  &  0.271057 - 0.677139$i$ \\
 
 1.0, \, 0.1 \, -0.0013 & 0.347633 - 0.121233$i$ & 0.310958 - 0.382017$i$ & 0.271118 - 0.677437$i$  \\
 
 1.0, \, 0.1 \, -0.0014 & 0.347784 - 0.121260$i$ & 0.311051 - 0.382156$i$ & 0.271178 - 0.677734$i$  \\
 
 1.0, \, 0.1 \, -0.0015 & 0.347934 - 0.121288$i$ & 0.311144 - 0.382295$i$ & 0.271235 - 0.678034$i$  \\
 
 1.0, \, 0.1 \, -0.0016 & 0.348085 - 0.121315$i$ & 0.311236 - 0.382434$i$ & 0.271290 - 0.678336$i$ \\
 
 1.0, \, 0.1 \, -0.0017 & 0.348235 - 0.121342$i$ & 0.311328 - 0.382573$i$ & 0.271344 - 0.678638 \\
 
 1.0, \, 0.1 \, -0.0018 & 0.348386 - 0.121369$i$ & 0.311420 - 0.382712$i$ & 0.271395 - 0.678941$i$  \\
  
 1.0, \, 0.1 \, -0.0019  & 0.348536 - 0.121396$i$ & 0.311511 - 0.382850$i$ & 0.271443 - 0.679247$i$ \\

 1.0, \, 0.1 \, -0.0020  & 0.348687 - 0.121422$i$ & 0.311603 - 0.382989$i$ & 0.271490 - 0.679552$i$ \\
   [0.2ex] 
 \hline \hline \hline 
\end{tabular}
\end{center}
\end{table}

\begin{table}[!h]
\begin{center}
\caption{\label{l2}Utilizing the sixth--order WKB approximation, we demonstrate the quasinormal frequencies corresponding to varying values of \( M \), \( \lambda \), and $\Lambda$ specifically for \( l=2 \).}
\begin{tabular}{c| c | c | c} 
 \hline\hline\hline 
 \!\!\!\! $M$ \,\,\,\,\,\,  $\lambda$  \,\,\,\, \,\, $\Lambda$  & $\omega_{0}$ & $\omega_{1}$ & $\omega_{2}$  \\ [0.2ex] 
 \hline 
 1.00, \, 0.10, \, -0.01 & 0.590234 - 0.123075$i$ & 0.561064 - 0.378474$i$ & 0.513965 - 0.656422$i$  \\ 
 
 1.01, \, 0.10, \, -0.01 & 0.584825 - 0.121924$i$ & 0.555866 - 0.374973$i$ & 0.509103 - 0.650431$i$  \\
 
 1.02, \, 0.10, \, -0.01 & 0.579527 - 0.120797$i$ & 0.550774 - 0.371544$i$ &  0.504335 - 0.644563$i$  \\
 
 1.03, \, 0.10, \, -0.01 & 0.574335 - 0.119691$i$ & 0.545783 - 0.368183$i$ &  0.499658 - 0.638814$i$  \\
 
 1.04, \, 0.10, \, -0.01 & 0.569248 - 0.118608$i$ & 0.540891 - 0.364889$i$ & 0.495071 - 0.633182$i$ \\
 
 1.05, \, 0.10, \, -0.01  & 0.564261 - 0.117546$i$ & 0.536095 - 0.361659$i$ & 0.490571 - 0.627663$i$  \\
 
 1.06, \, 0.10, \, -0.01 & 0.559372 - 0.116504$i$ & 0.531393 - 0.358493$i$ & 0.486154 - 0.622253$i$ \\
 
 1.07, \, 0.10, \, -0.01 & 0.554578 - 0.115482$i$ & 0.526781 - 0.355388$i$ & 0.481818 - 0.616951$i$  \\
 
 1.08, \, 0.10, \, -0.01  & 0.549877 - 0.114480$i$ & 0.522258 - 0.352343$i$ & 0.477561 - 0.611752$i$ \\
  
 1.09, \, 0.10, \, -0.01 & 0.545265 - 0.113497$i$ & 0.517820 - 0.349356$i$ & 0.473380 - 0.606654$i$  \\

  1.10, \, 0.10, \, -0.01 & 0.540742 - 0.112532$i$ & 0.513465 - 0.346425$i$ & 0.469274 - 0.601655$i$ \\
   [0.2ex] 
 \hline \hline \hline 
  \!\!\!\! $M$ \,\,\,\,\,\,  $\lambda$  \,\,\,\, \,\, $\Lambda$   & $\omega_{0}$ & $\omega_{1}$ & $\omega_{2}$  \\ [0.2ex] 
 \hline 
 1.00, \, 0.10,  \, -0.01 & 0.590234 - 0.123075$i$ & 0.561064 - 0.378474$i$ & 0.513965 - 0.656422$i$ \\ 
 
 1.00, \, 0.11, \, -0.01 & 0.599921 - 0.125781$i$ & 0.570043 - 0.386850$i$ & 0.521905 - 0.671080$i$ \\
 
 1.00, \, 0.12, \, -0.01 & 0.609892 - 0.128579$i$ & 0.579278 - 0.395516$i$  &  0.530064 - 0.686250$i$  \\
 
 1.00, \, 0.13, \, -0.01 & 0.620157 - 0.131475$i$ & 0.588780 - 0.404485$i$ & 0.538451 - 0.701957$i$  \\
 
 1.00, \, 0.14, \, -0.01 & 0.630729 - 0.134473$i$ & 0.598559 - 0.413773$i$ & 0.547076 - 0.718226$i$ \\
 
 1.00, \, 0.15, \, -0.01 & 0.641622 - 0.137578$i$ & 0.608627 - 0.423394$i$ & 0.555949 - 0.735083$i$  \\
 
 1.00, \, 0.16, \, -0.01 & 0.652848 - 0.140794$i$ & 0.618996 - 0.433365$i$ & 0.565079 - 0.752559$i$ \\
 
 1.00, \, 0.17, \, -0.01 & 0.664421 - 0.144128$i$ & 0.629679 - 0.443702$i$ & 0.574477 - 0.770682$i$ \\
 
 1.00, \, 0.18, \, -0.01 & 0.676359 - 0.147585$i$ & 0.640689 - 0.454423$i$ & 0.584154 - 0.789487$i$  \\
  
 1.00, \, 0.19, \, -0.01 & 0.688675 - 0.151172$i$ & 0.652040 - 0.465549$i$ & 0.594123 - 0.809006$i$  \\

  1.00, \, 0.20, \, -0.01 & 0.701387 - 0.154894$i$ & 0.663748 - 0.477100$i$ & 0.604395 - 0.829277$i$  \\
   [0.2ex] 
 \hline \hline \hline 
  \!\!\!\! $M$ \,\,\,\,\,\,  $\lambda$  \,\,\,\, \,\, $\Lambda$  & $\omega_{0}$ & $\omega_{1}$ & $\omega_{2}$  \\ [0.2ex] 
 \hline 
 1.0, \, 0.1 \, -0.0010 & 0.570273 - 0.119902$i$ & 0.544435 - 0.367128$i$ & 0.501786 - 0.633691$i$ \\ 
 
 1.0, \, 0.1 \, -0.0011 & 0.570498 - 0.119939$i$ & 0.544624 - 0.367256$i$ & 0.501935 - 0.633944$i$ \\
 
 1.0, \, 0.1 \, -0.0012 & 0.570722 - 0.119975$i$ & 0.544813 - 0.367384$i$ &  0.502085 - 0.634196$i$ \\
 
 1.0, \, 0.1 \, -0.0013 & 0.570947 - 0.120011$i$ & 0.545002 - 0.367512$i$ & 0.502233 - 0.634449$i$  \\
 
 1.0, \, 0.1 \, -0.0014 & 0.571172 - 0.120048$i$ & 0.545191 - 0.367640$i$ & 0.502382 - 0.634701$i$  \\
 
 1.0, \, 0.1 \, -0.0015 & 0.571396 - 0.120084$i$ & 0.545380 - 0.367768$i$ & 0.502530 - 0.634954$i$  \\
 
 1.0, \, 0.1 \, -0.0016 & 0.571621 - 0.120121$i$ & 0.545569 - 0.367896$i$ & 0.502679 - 0.635207$i$  \\
 
 1.0, \, 0.1 \, -0.0017 & 0.571845 - 0.120157$i$ & 0.545758 - 0.368024$i$ & 0.502827 - 0.635459$i$ \\
 
 1.0, \, 0.1 \, -0.0018 & 0.572069 - 0.120193$i$ & 0.545946 - 0.368152$i$ & 0.502975 - 0.635711$i$  \\
  
 1.0, \, 0.1 \, -0.0019  & 0.572293 - 0.120229$i$ & 0.546135 - 0.368280$i$ & 0.503122 - 0.635964$i$ \\

 1.0, \, 0.1 \, -0.0020  & 0.572518 - 0.120266$i$  &  0.546323 - 0.368407$i$ & 0.503269 - 0.636216$i$ \\
   [0.2ex] 
 \hline \hline \hline 
\end{tabular}
\label{quasinormalmodestable3}
\end{center}
\end{table}


\section{Emission rate}
	
The energy emission rate can be obtained according to \cite{mustafa2022shadows}
\begin{equation}\label{emission}
	\frac{{{\mathrm{d}^2}E}}{{\mathrm{d}\omega \mathrm{d}t}} = \frac{{2{\pi ^2}\sigma_{lim} {\omega ^3}}}{{{e^{\frac{\omega }{T_{H}}}} - 1}},
\end{equation}
where $\omega$ represents the frequency of photon, and $T$ is the Hawking temperature for the outer event horizon, being defined as
\begin{equation}
	T_{H} = \left.\frac{f'(r)}{4\pi}\right|_{r=r_{h}},
\end{equation}
where 
 $f(r)$ is taken in accordance with Eq. \eqref{LapseF} and $r_{h}$ represents the radius of the horizon and $\sigma_{lim}$ is the limiting cross section intricately connected to the radius of the event horizon. The equation governing $\sigma_{lim}$ in $d$ spacetime dimensions is provided in Reference \cite{123}, and additional insights can be found in the works of \cite{124,125}
\begin{equation}
	\sigma_{lim}=\frac{\pi^{\frac{d-2}{2}}R_{s}^{d-2}}{\Gamma\left(\frac{d}{2}\right)}, 
\end{equation}
where $R_{s}$ represents the shadow radius and $\Gamma\left(\frac{d}{2}\right)$ denotes the gamma function.

To ascertain the absorption cross--section, we have two methodologies at our disposal. The initial approach involves deriving the effective potential using Equation \eqref{effpot}. Meanwhile, the second method entails incorporating the purely imaginary quantity represented as $\Gamma$, thereby facilitating the analysis
\begin{equation}
	\Gamma=\frac{i\left(\tilde{\omega}^{2}-V_{0}\right)}{\sqrt{-2V''_{0}}}-\sum_{j=2}^{6} \Lambda_{j},
\end{equation}
where $V_{0}$ is the potential calculated in its maximum point. The aforementioned equation stems from the WKB method. Within this framework, the coefficients corresponding to the complex functions defining the effective potential are denoted as $\Lambda_{j}$, and $\tilde{\omega}$ signifies a purely real frequency associated with quasinormal modes. The reflection and transmission coefficients are articulated as follows:
\begin{equation}
	\left|R\right|^{2} = \frac{1}{1+e^{-2i\pi \gamma}},\qquad \left|T\right|^{2} = \frac{1}{1+e^{2i\pi \gamma}},
\end{equation} 
The partial absorption cross--section for $\ell$ modes is precisely defined as:
\begin{equation}
	\sigma_{\ell} = \frac{\pi \left(2\ell +1\right)}{\tilde{\omega}^{2}} \left|T_{\ell}\left(\tilde{\omega}\right)\right|^{2}.
\end{equation}  
The total absorption cross--section is established through the summation of all partial absorption cross--sections, as expressed by:
\begin{equation}
	\sigma_{abs} = \sum_{\ell}^{} \sigma_{\ell}.
\end{equation} 
Moreover, by using Eq. \eqref{eq:rshnotasymptoticallyflat} for $d=2$ it can be written as 
\begin{equation}
\sigma_{lim} = \pi R_{s}^{2} = \pi \left[\frac{3 (1-\lambda) M \sqrt{\frac{-6 \lambda  M+6 M+\Lambda  r_{O}^3-3 r_{O}}{(\lambda -1) r_{O}}}}{\sqrt{\frac{1}{1-\lambda }+9 (\lambda -1) \Lambda  M^2}}\right]^{2}.
\end{equation}

\begin{figure}[htbp]
	\centerline{
		\includegraphics[scale = 0.5]{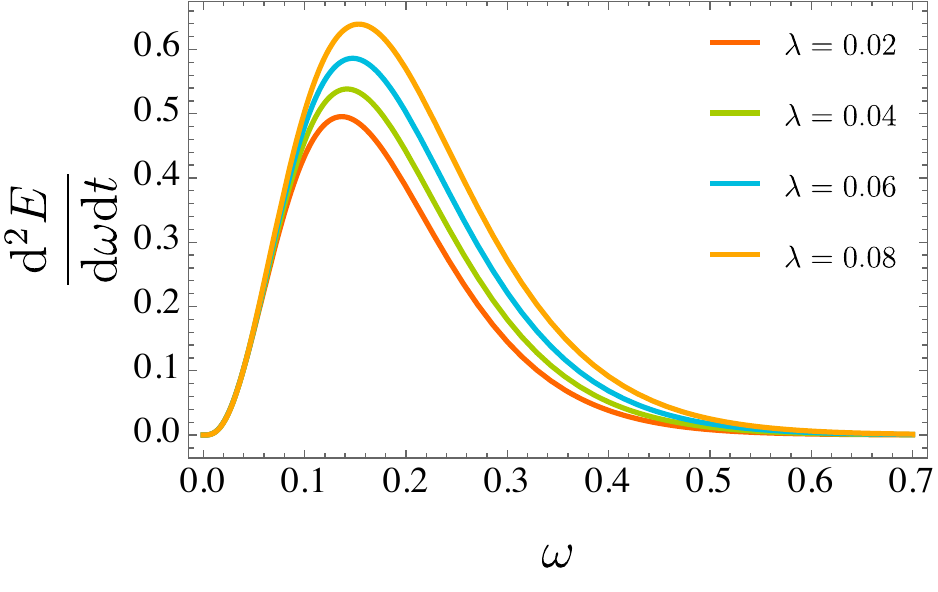}\hspace{0.5cm}
		\includegraphics[scale = 0.5]{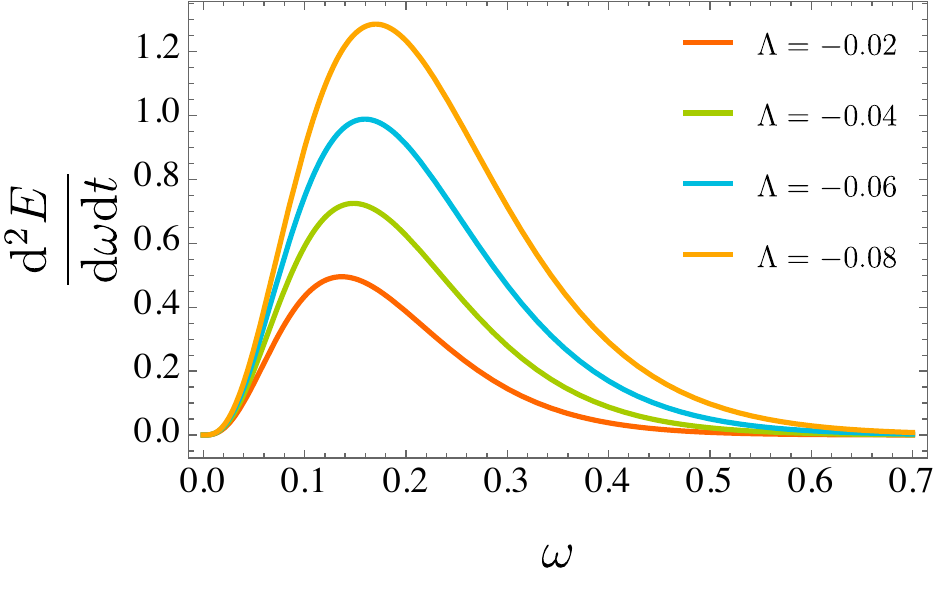}} \vspace{-0.2cm}
	\caption{The emission rates are represented with respect to
		frequency for $M=1$, in left plot different values of $\lambda$, fixed values of $\Lambda=-0.02$  and in right plot different values of $\Lambda$ and fixed values of $\lambda=0.02$  are considered.}
	\label{fig:emission}
\end{figure}

The emission rate as a function of frequency $\omega$ is depicted in Fig. \ref{fig:emission}. In the left panel, the $\Lambda$ value is held constant, while different values of $\lambda$ (specifically, $0.02, 0.04, 0.06, 0.08$) are considered. Conversely, in the right panel, $\lambda$ is fixed, and varying $\Lambda$ values of $-0.02, -0.04, -0.06, -0.08$ are explored.

In the left panel, discernible peaks in the emission rate are observed, indicating that the maximum energy emission increases with higher values of the parameter $\lambda$. Additionally, these peaks shift towards higher frequencies. Conversely, in the right panel, as the absolute value of $\Lambda$ increases, the emission rate's peak shifts to higher frequencies, accompanied by an increase in the maximum emission value.



\section{Conclusion}

This study explored the impact of antisymmetric tensor effects on spherically symmetric black holes, examining photon spheres, shadows, emission rate, and quasinormal frequencies concerning a parameter that triggered Lorentz symmetry breaking. Both scenarios, one without and one with the presence of a cosmological constant, were investigated. In the first scenario, the Lorentz violation parameter, denoted as $\lambda$, played a pivotal role in reducing both the photon sphere and the shadow radius, while also inducing a damping effect on quasinormal frequencies. Conversely, in the second scenario, as the values of the cosmological constant ($\Lambda$) increased, an expansion in the shadow radius was observed, accompanied by a reduction in damping modes associated with gravitational wave frequencies. Furthermore, we determined the shadow constraints by analyzing observational data acquired from the Event Horizon Telescope (EHT), specifically concentrating on the shadow images of Sagittarius $A^{*}$.


\section*{Acknowledgments}
\hspace{0.5cm}

The authors would like to thank the anonymous referee for a careful reading of the manuscript and for the remarkable suggestions given to us. A. A. Araújo Filho would like to thank Fundação de Apoio à Pesquisa do Estado da Paraíba (FAPESQ) and Conselho Nacional de Desenvolvimento Cientíıfico e Tecnológico (CNPq)  -- [150891/2023-7] for the financial support. Most of the calculations were performed by using the \textit{Mathematica} software. Also, The research was supported by the Long--Term Conceptual Development of a University of Hradec Kralove for 2023, issued by the Ministry of Education, Youth, and Sports of the Czech Republic.


\bibliographystyle{ieeetr}

\bibliography{main}

\end{document}